\documentclass[useAMS,usenatbib,usegraphicx]{mn2e}

\usepackage{amsmath,amssymb,amsbsy,amsfonts}
\usepackage[british]{babel}
\usepackage{mdwmath}
\usepackage{color}
\usepackage{url}
\usepackage{array}
\usepackage{graphicx}
\usepackage{units}
\usepackage{hyperref}
\usepackage{times}
\usepackage{pdflscape}
\usepackage{balance}
\usepackage{stackengine}
\usepackage{scalerel}
\usepackage{textcomp}
\usepackage{soul}
\usepackage{cancel}

\definecolor{custom-blue}{RGB}{41,22,206}
\hypersetup{
  colorlinks   = true, 
  urlcolor     = blue, 
  linkcolor    = blue, 
  citecolor    = blue  
}

\usepackage[abs]{overpic}
\usepackage{xcolor,varwidth}

\usepackage[scr=boondoxo,scrscaled=1.05]{mathalfa}
\definecolor{dodgerblue}{RGB}{30,144,255}

\newcommand{\eq}[1]{equation~\eqref{#1}}
\newcommand{\eqs}[1]{equations~\eqref{#1}}

\newcommand{\co}{\mathrm{const.}}
\newcommand{\ud}{\mathrm{d}}

\newcommand{\s}[1]{{\text{\tiny $#1$ }}\hspace{-2pt}}

\newcommand{\8}{\s{\infty}}

\renewcommand{\theequation}{\thesection.\arabic{equation}}

\title[Spherical accretion onto rotating black holes]{
Spherical accretion: Bondi, Michel, and rotating black holes
}

\author[A.~Aguayo-Ortiz et al.]{
Alejandro Aguayo-Ortiz$,^1$\thanks{E-mail: aaguayo@astro.unam.mx,
emilio.tejeda@conacyt.mx, olivier.sarbach@umich.mx, diego@astro.unam.mx}
Emilio Tejeda,$^{2}$ Olivier Sarbach,$^3$ \& Diego L\'opez-C\'amara$^4$ \\
$^1$ Universidad Nacional Aut\'onoma de M\'exico, Instituto de Astronom\'ia, AP 70-264, CDMX 04510, M\'exico \\ 
$^2$ C\'atedras CONACyT --
Instituto de F\'isica y Matem\'aticas, Universidad Michoacana de San Nicol\'as
de Hidalgo, Edificio C-3, Ciudad Universitaria, 58040 Morelia,\\
\hspace{0.15cm} Michoac\'an, Mexico \\
$^3$ Instituto de F\'isica y Matem\'aticas, Universidad Michoacana 
de San Nicol\'as de Hidalgo, Edificio C-3, Ciudad Universitaria, 58040 
Morelia, Michoac\'an, Mexico \\
$^4$ C\'atedras CONACyT -- Universidad Nacional Aut\'onoma de M\'exico, Instituto de Astronom\'ia, AP 70-264, CDMX 04510, M\'exico}

\pagerange{\pageref{firstpage}--\pageref{lastpage}}

\begin{document}

\label{firstpage}

\maketitle

\begin{abstract}
In this work we revisit the steady state, spherically symmetric gas accretion
problem from the non-relativistic regime to the ultra-relativistic one. We
first perform a detailed comparison between the Bondi and Michel models, and
show how the mass accretion rate in the Michel solution approaches a constant
value as the fluid temperature increases, whereas the corresponding Bondi
value continually decreases, the difference between these two predicted values
becoming arbitrarily large at ultra-relativistic temperatures. Additionally,
we extend the Michel solution to the case of a fluid with an equation of
state corresponding to a monoatomic, relativistic gas. Finally, using general
relativistic hydrodynamic simulations, we study spherical accretion onto a
rotating black hole, exploring the influence of the black hole spin on the
mass accretion rate, the flow morphology and characteristics, and the sonic
surface. The effect of the black hole spin becomes more significant as the gas
temperature increases and as the adiabatic index $\gamma$ stiffens. For an
ideal gas in the ultra-relativistic limit ($\gamma=4/3$), we find a reduction
of 10 per cent in the mass accretion rate for a maximally rotating black
hole as compared to a non-rotating one, while this reduction is of up to 50
per cent for a stiff fluid ($\gamma=2$).
\end{abstract}


\section{Introduction}
\label{sec:introduction}

Gas accretion onto a compact gravitating object is one of the most studied
problems in astrophysics. In one of the pioneering works of accretion theory,
\citet{bondi1952} found an analytic solution for the spherically symmetric,
steady-state accretion flow of an infinite gas cloud onto a Newtonian
point-mass potential. This model has been widely extended and applied in
many different fields in astrophysics, from the study of star formation
to cosmology. See~\citet{armitage2020} for a recent historical review of
this subject.

One of the first studies of spherical accretion onto black holes was the
extension of the Bondi solution into the general relativistic regime performed
by~\citet{michel1972}. In this study, Michel found an analytic solution
describing the spherical accretion of a polytropic gas onto a Schwarzschild
black hole. A formal mathematical analysis of the Michel solution for a generic
equation of state (EoS) can be found in~\citet{chaverra2016b}. Following
Michel's procedure, other authors have found semi-analytic generalizations
for different types of spherically-symmetric (non-rotating) black hole
solutions~\citep[e.g.][]{chaverra2015, miller2017,yang2020,abbas2021}. In
a recent work, \citet{richards2021a} explore the non-relativistic and
ultra-relativistic limits of Michel's solution, mainly focusing on a gas
with a stiff EoS (values of the adiabatic index larger than 5/3).

In past decades, the spherical accretion Bondi model has been revisited
and extended, by including different additional physical ingredients. For
example, some authors have taken into account the fluid's self gravity by
solving the coupled Einstein-Euler system in spherical symmetry, either
with an analytical treatment~\citep{malec1999} or by performing numerical
simulations~\citep{lora2013a}. Some works have considered the extension
of a Bondi-like solution by introducing a low angular momentum fluid
~\citep{abramowicz1981,proga2003,mach2018}, finding a transition between
a quasi-spherical accretion flow and the formation of a thick torus in
the equatorial plane. Similarly, there have been works studying spherical
accretion in the presence of magnetic fields, either assuming a central
dipole~\citep{toropin1999}, or by including a three-dimensional, large-scale
weak magnetic field~\citep[e.g.][]{igumenshchev2002,ressler2021}. Together with
magnetic fields, some works have included the effects of a radiation field,
addressing the problem either with a simplified approach~\citep[][where
the author considered a radiation-dominated fluid]{begelman1978} or
with a self-consistent, radiative-transfer treatment using numerical
simulations~\citep{mckinney2014,weih2020}. In this regard, there have also
been studies that extract the shadow of the spherically accreted, optically
thin cloud around a non-rotating black hole~\citep{narayan2019}. Other
studies have considered the effects of thermal conduction on magnetized
spherical accretion flows~\citep{sharma2008}, vorticity~\citep{krumholz2005},
or studied the spherical accretion of a relativistic collisionless kinetic
(i.e.~a Vlasov) gas~\citep[][]{rioseco2017a}.

Recent works have also studied deviations away from spherical symmetry
by introducing large-scale, small-amplitude density anisotropies, finding
that even a slight equator-to-poles density contrast can drastically modify
Bondi's solution, giving rise to an inflow-outflow configuration consisting
of equatorial accretion and a bipolar outflow. The resulting steady-state
configuration, dubbed choked accretion, was studied in \cite{ATH2019} at the
Newtonian level and, within a general relativistic framework, in \cite{TAH2020}
and \cite{AST2021} for Schwarzschild and Kerr black holes, respectively. In
these series of works, it was found that the total mass flux that reaches
the central accretor is of the order of magnitude of the corresponding Bondi
mass accretion rate, while all the excess flux is redirected by the density
gradient as outflow. Under the conditions explored so far, the Bondi mass
accretion rate acts as a threshold value delimiting whether a given incoming
flow becomes choked and prone to the ejection of a bipolar outflow.

Among the astrophysical applications of the spherical
accretion model, we mention the study of gas accretion in an
expanding Universe~\citep[][]{colpi1996}, the formation and
growth of primordial black holes in the early stages of the
Universe~\citep{zeldovich1967,carr1981,karkowski2013,lora2013a}, and
the accretion onto a mini black hole from the interior of a neutron star
\citep{kouvaris2014,genolini2020,richards2021b}. On the other hand, the Bondi
solution allows to estimate, by providing useful characteristic scale tools,
the accretion and growth rate of the central supermassive black hole at the
centre of galaxies~\citep{maraschi1974,moscibrodzka2006,ciotti2017,moffat2020}
and active galactic nuclei~\citep{krolik1983,russel2013,russel2015}, where
observations provide information only from regions far away from the central
accretor. Similarly, the Bondi prescription is often used in cosmological
simulations as a sub-grid model to estimate the accretion rate of gas onto
supermassive black holes at galactic centres~\citep{dave2019}.

On the other hand, the analytic study of accretion flows onto rotating black
holes has proven more challenging. Notably, \citet*{petrich1988} found a
full analytic solution that describes the accretion of an irrotational,
ultra-relativistic stiff fluid onto a rotating Kerr black hole. However,
a main caveat of this solution is that it requires a rather specific,
unphysical EoS, in which the sound speed equals the speed of light.\footnote{An
ultra-relativistic stiff fluid corresponds to the relativistic generalisation
of an incompressible fluid in Newtonian hydrodynamics \citep{tejeda2018}.}
Assuming a more general EoS, \citet{beskin1995} studied the problem
of spherical accretion onto a slowly rotating black hole by means of a
perturbative analysis, and~\citet{pariev1996} extended this work to the case of
a rapidly rotating black hole. Both \citet{beskin1995} and \citet{pariev1996}
considered only small deviations away from a Bondi background solution, in
other words, these studies where limited to the case of non-relativistic
values for the gas temperature at infinity. Even though this assumption
might be reasonable in many astrophysical settings, the determination of
the effect of the black hole spin on the accretion flow given an arbitrary
gas temperature remains an open problem.

The applications of Bondi's model in most of the aforementioned works consider
the gas accretion in the non-relativistic regime, not to mention that they
neglect the rotation of the black hole. The reason for this is that the Bondi
scale factors are estimated and measured at distances far away from the central
black hole, where it is safe to neglect relativistic effects. Nevertheless,
in order to analyse the exact differences between the Bondi solution and the
relativistic extension performed by Michel, as well as to assess the effect
of the black hole spin, it is important to perform a quantitative study of
the consequences of having relativistic gas temperatures and strong gravity
fields in the vicinity of a rotating black hole.

In this work we study the spherically symmetric gas accretion problem from
the non-relativistic regime to the ultra-relativistic one, considering both
rotating and non-rotating black holes.\footnote{By `spherically symmetric'
accretion problem onto a rotating black hole, we refer to the gas state
being spherically symmetric asymptotically far away from the central
black hole. Clearly, a rotating black hole does not admit a spherically
symmetric solution at finite radii.} We first perform a detailed comparison
between the~\citet{bondi1952} and~\citet{michel1972} models by studying the
behaviour of the relativistic solution across a wide range of values of the
gas temperature. In particular, we discuss in detail the isothermal, the
non-relativistic, and the ultra-relativistic limits of the Michel solution.
We then extend Michel's solution to the case of a monoatomic gas obeying
a relativistic EoS~\citep{juttner1911,taub1948,synge1957}. We also revisit
\citet{petrich1988}'s analytic solution and apply it to the particular case
of a spherically symmetric accretion flow onto a Kerr black hole. Finally, by
means of two dimensional (2D) general relativistic hydrodynamic simulations,
we perform a quantitative study of the effect that the black hole spin has
on the spherical accretion problem, focusing in particular on its effects
on the mass accretion rate and on the flow morphology for several EoS. As
part of this study, we show how, under the appropriate limits, the obtained
numerical results coincide with the analytic solutions of \cite{michel1972}
and \citet{petrich1988}.

The paper is organised as follows. In Section~\ref{sec:analytic} we
discuss the analytic solutions of \citet{bondi1952}, \citet{michel1972} and
\citet{petrich1988}. In Section~\ref{sec:polytropic} we present our numerical
study of the spherical accretion of a perfect fluid onto a rotating black
hole. Finally, in Section~\ref{sec:summary} we present a summary of the main
results found in this article and give our conclusions. Technical details
regarding the isothermal and non-relativistic limits of the Michel solution,
the correct determination of the sonic surface for flows on rotating black
holes, and orthonormal frames are discussed in appendices.

\setcounter{equation}{0}
\section{Analytic solutions}
\label{sec:analytic}

In this section we review three analytic solutions describing a steady-state,
spherical accretion flow onto a central massive object. We start by revisiting
the~\citet{bondi1952} solution and perform a detailed comparison with the
relativistic extension found by~\citet{michel1972}. Then, we extend the
latter solution by considering the more realistic equation of state for a
monoatomic relativistic gas introduced by~\citet{juttner1911}. Finally, in
order to give a description of accretion onto a rotating black hole, we also
discuss the ultra-relativistic, stiff solution found by~\citet{petrich1988}
in the case of spherical symmetry.

\subsection{Bondi solution}
\label{SubSec:Bondi}

In the~\citet{bondi1952} analytic solution, one considers an infinite,
spherically symmetric gas cloud accreting onto a Newtonian central object
of mass $M$. At large distances, the gas cloud is assumed to be at rest and
characterised by a homogeneous density $\rho_\8$ and pressure $P_\8$. Note
that, using an ideal gas EoS, we can alternatively describe the state of
the fluid in terms of the dimensionless gas temperature $\Theta$ defined as:
\begin{equation}
   \Theta = \frac{k_{\rm B}\,T}{\bar{m}\, c^2} = \frac{P}{\rho\, c^2},
\end{equation} 
where $c$ is the speed of light, $k_{\rm B}$ Boltzmann's constant, and
$\bar{m}$ the average rest mass of the gas particles. As reference values,
$\Theta \simeq T/(10^{13}\,{\rm K})$ for atomic hydrogen gas and $\Theta
\simeq T/(10^{10}\,{\rm K})$ for an electron-positron plasma.

Under the assumptions of steady-state and spherical symmetry, the equations
governing the Bondi accretion flow are the continuity equation and the radial
Euler equation, i.e.
\begin{subequations}
   \begin{align}
      \frac{1}{r^2}\frac{\ud }{\ud r}\left(r^2\rho\,v\right) = 0, \label{eb.1}\\
      v\frac{\ud v}{\ud r}+\frac{1}{\rho}\frac{\ud P}{\ud r} +\frac{GM}{r^2}=
      0, \label{eb.2}
\end{align}
\end{subequations}
where $v=|\ud r/\ud t|$ is the radial velocity of the fluid.

Considering that, in addition to the ideal gas EoS, the fluid obeys
a polytropic relation $P=K\rho^\gamma$, with $K=\co$ and $\gamma$ the
adiabatic index (assumed to lie in the range $1\le\gamma\le2$), \eqs{eb.1}
and \eqref{eb.2} can be integrated:
\begin{subequations}
   \begin{align}
      4\pi\,r^2\rho\,v=\dot{M}&=\co, 
      \label{eb.3}\\
      \frac{v^2}{2}+ \mathscr{h} -\frac{GM}{r}= \mathscr{h}\8 &= \co,
      \label{eb.4}
   \end{align}
\end{subequations}
where
\begin{equation} 
   \mathscr{h} = \left(\frac{\gamma}{\gamma-1}\right)\frac{P}{\rho} =
   \frac{\gamma\,\Theta\,c^2}{\gamma - 1} = \frac{\mathcal{C}^2}{\gamma - 1}
   \label{eb.5}
\end{equation}
is the specific enthalpy and $\mathcal{C} := \sqrt{\partial P/\partial
\rho}$ the adiabatic speed of sound. Note that \eq{eb.5} is only valid for
$\gamma>1$. In the isothermal case, where $\gamma=1$ and $\Theta\equiv\Theta\8=
\mathcal{C}^2_\8/c^2$, \eq{eb.5} needs to be replaced by
\begin{equation}
   \mathscr{h} - \mathscr{h}_\8=\mathcal{C}^2_\8
   \ln\left(\frac{\rho}{\rho\8}\right).
   \label{eb.6}
\end{equation}

In addition to the steady-state and spherical symmetry conditions, Bondi also
assumed that the flow is transonic, i.e.~that there exists a radius $r_s$
at which the fluid's radial velocity equals the local speed of sound. From
\eqs{eb.1} and \eqref{eb.2}, it is simple to calculate that the fluid at
the sonic radius, $r_s$, satisfies
\begin{subequations}
   \begin{gather}
        r_s = \frac{GM}{2v_s^2},\\
        v_s = \mathcal{C}_s = \mathcal{C}_\8 \left(\frac{2}{5-3\gamma}
        \right)^{1/2}.
   \end{gather}
\end{subequations}

The transonic solution found by Bondi is unique and maximises the accretion
rate onto the central object, which, in turn, is given by
\begin{equation}
   \dot{M}_{\rm B} = 4\pi\,\lambda_{\rm B} (G M)^2
   \frac{\rho_\8}{\mathcal{C}^{3}_\8},
   \label{bm}
\end{equation}
where $\lambda_{\rm B}$ is a numerical factor of order one that depends only
on $\gamma$ and is given by
\begin{equation}
   \lambda_{\rm B} = \frac{1}{4}
   \left(\frac{2}{5-3\gamma}\right)^{\frac{5-3\gamma}{2(\gamma-1)}}.
   \label{lm}
\end{equation}
The accretion rate given in \eq{bm} is only valid for \mbox{$\gamma \le
5/3$}. In order to discuss the \mbox{$\gamma> 5/3$} case, one must
necessarily account for general relativistic effects as we shall see in
Section~\ref{subsec:michel}. Particular values for $\lambda_{\rm B}$ in
\eq{lm} are
\begin{align}
   &\lambda_{\rm B}(5/3)=1/4,
   \nonumber \\
   &\lambda_{\rm B}(4/3)=1/\sqrt{2} \simeq 0.71,
   \nonumber \\
   &\lambda_{\rm B}(1)=\mathrm{e}^{3/2}/4 \simeq 1.12.
   \nonumber
\end{align}

\begin{figure}
   \centering
   \includegraphics[width=0.475\textwidth]{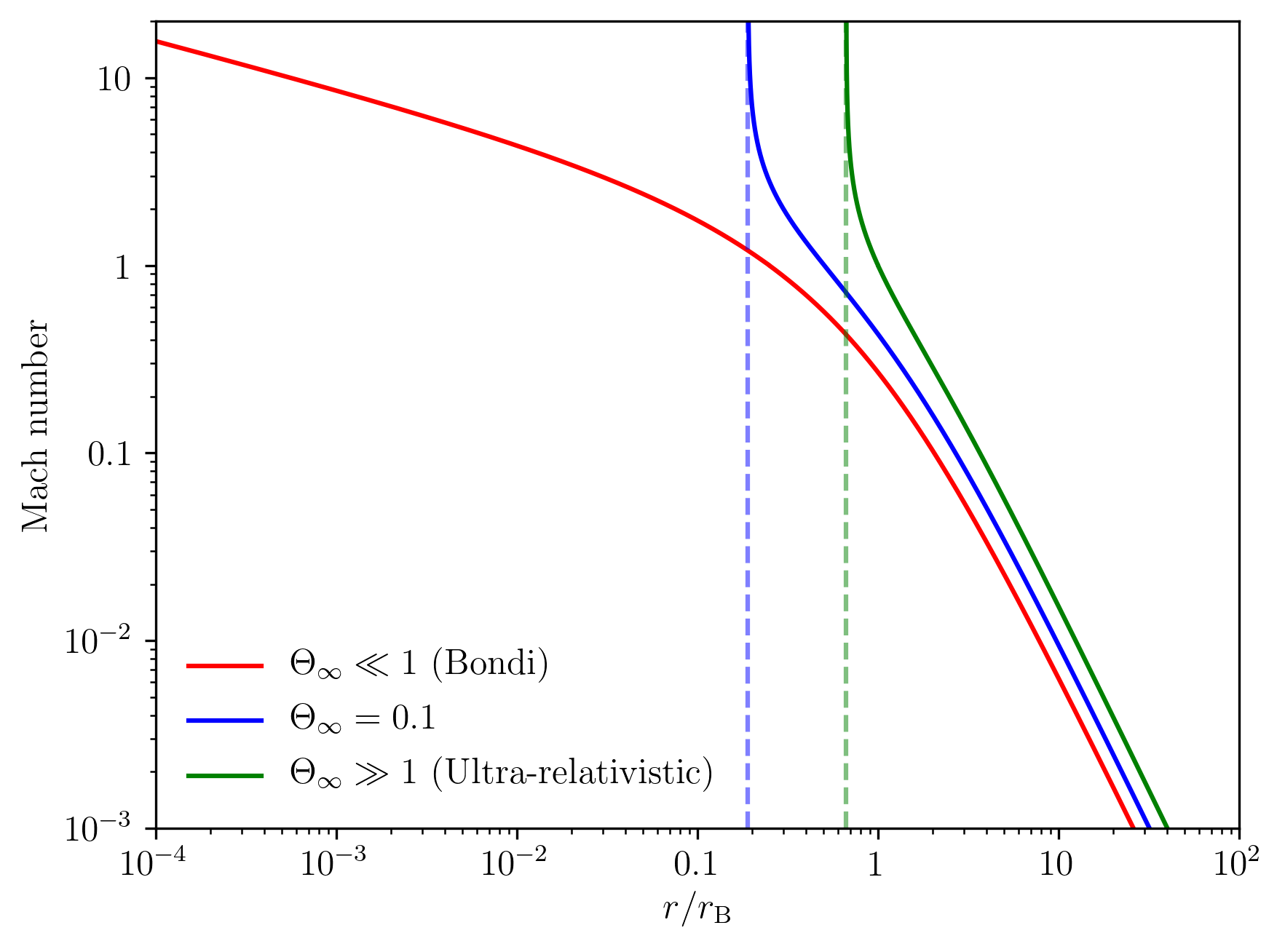}
   \caption{Mach number as a function of radius for the case of a
   \mbox{$\gamma=4/3$} polytrope and for three asymptotic temperatures. Note
   that the non-relativistic limit ($\Theta_\8\ll1$) corresponds to the Bondi
   solution. The vertical dashed lines show the location of the black
   hole's event horizon for $\Theta_\8 = 0.1$ and $\Theta_\8 \gg 1$. In
   all cases the accretion flow has transitioned from subsonic to supersonic
   before crossing the event horizon. The horizontal axis is scaled in units
   of the Bondi radius $r_\mathrm{B}=GM/C^2_\8$.}
   \label{fig:mach}
\end{figure}

An interesting characteristic of the Bondi solution is that it can be
written in a scale-free form with respect to the mass of the central
object $M$ and the thermodynamic state of the fluid \mbox{($\rho_\8$, $P_\8$)}
by adopting $r_\mathrm{B}=GM/C^2_\8$, $\mathcal{C}_\8$, and $\rho_\8$
as units of length, velocity, and density, respectively. In other words,
a global solution of the Bondi accretion problem is fully characterised
once a given value for the adiabatic index $\gamma$ is provided. Once the
value for the mass accretion rate of Bondi's solution for a given $\gamma$
is known, one can go back to \eqs{eb.3}--\eqref{eb.5} and solve numerically
the corresponding algebraic system of non-linear equations to obtain $\rho$,
$P$, and $v$ as a function of radius. See Figure~\ref{fig:mach}, for an
example where we show the resulting Mach number ($\mathcal{M}=v/\mathcal{C}$)
as a function of radius, for the solution with $\gamma=4/3$ (red line).

\subsection{Michel solution}
\label{subsec:michel}

As mentioned in the introduction, a general relativistic extension of the Bondi
solution was presented by \cite{michel1972} who considered a Schwarzschild
black hole as central accretor. In what follows we review Michel's solution
and discuss its main differences with respect to the Bondi model. It is
important to remark that the Michel solution assumes an ideal gas EoS that
follows a polytropic relation $P=K\,\rho^\gamma$, where, as in the
previous section, $\rho$ is the rest-mass density. Note however that this
assumption is limited in general. For example, for a monoatomic ideal gas,
it is only valid at non-relativistic temperatures (for which $\gamma=5/3$),
or at ultra-relativistic temperatures (for which $\gamma=4/3$). In order
to study the whole temperature domain in a consistent way, the polytropic
restriction must be dropped and a relativistic EoS \citep[as derived, for
example, from relativistic kinetic theory,][]{synge1957} must be adopted. We
discuss the extension of the Michel solution to such a relativistic EoS in
Section~\ref{subsec:releos}.

As in the Newtonian case, the governing equations are the conservation of
mass and energy, i.e.~the continuity equation and the requirement for the
energy-momentum tensor to be divergence-free,
\begin{subequations}
   \begin{align}
      \left(\rho\,U^\mu\right)_{;\mu} = 0, 
      \label{em.1}\\
      \left(T^{\mu\nu}\right)_{;\mu} = 0, 
      \label{em.2}
   \end{align}
\end{subequations}
where the semicolon stands for covariant derivative, $U^\mu$ is the fluid
four-velocity, $T^{\mu\nu} = \rho\,h\,U^\mu U^\nu + p\,g^{\mu\nu}$ is
the stress-energy tensor of a perfect fluid, $h=1+\mathscr{h}$ is the specific
relativistic enthalpy, and $g^{\mu\nu}$ denote the components of the inverse
of the Schwarzschild metric
\begin{equation}
ds^2 = -\left( 1 - \frac{2M}{r} \right) \ud t^2 +
   \frac{\ud r^2}{1 - \frac{2M}{r}}
  + r^2\left(\ud\theta^2 
   + \sin^2\theta\ud\phi^2\right).
\end{equation}
In order to ease the notation, we adopt geometrised units in which $G=c=1$.

It is useful to recall at this point that, in the relativistic regime,
the fluid's sound speed is defined as
\begin{equation}
   \mathcal{C}^2 := \frac{\rho}{h}\frac{\partial h}{\partial \rho} =
   \frac{\gamma}{h}\frac{P}{\rho} = \frac{\gamma}{h}\,\Theta,
   \label{em.5}
\end{equation}
where, for the second equal sign, we have substituted the polytropic relation
for a perfect fluid. Also note that \eq{em.5} can be recast to express $h$
in terms of $\mathcal{C}$ or $\Theta$ as
\begin{equation}
   h = \frac{1}{1-\mathcal{C}^2/(\gamma-1)} = 1 + \frac{\gamma}{\gamma
   -1}\Theta.
\end{equation}

Under the conditions of steady-state and spherical symmetry, \eqs{em.1}
and \eqref{em.2} reduce to
\begin{subequations}
   \begin{align}
      \frac{\ud}{\ud r}\left(r^2\rho\,U^r\right) = 0, 
      \label{em.1.2}\\
      \frac{\ud}{\ud r}\left(r^2\rho\,h\,U_t\,U^r\right) = 0, 
      \label{em.2.2}
   \end{align}
\end{subequations}
which, upon integration, can be rewritten as
\begin{subequations}
   \begin{align}
      4\pi\,r^2\rho\,u=\dot{M}=\co, 
      \label{em.3}\\
      h\left(1-\frac{2M}{r}+u^2\right)^{1/2}=h\8 = \co, 
      \label{em.4}
   \end{align}
\end{subequations}
where $u = |U^r|$. 

As in the Newtonian case, there exists a unique, transonic
solution where the fluid is at rest asymptotically far away from
the central object and that is regular across the black hole's event
horizon~\citep{chaverra2015,chaverra2016b}. In order to find the defining
conditions that are satisfied at the sonic point $r_s$, it is useful to combine
\eqs{em.1.2} and \eqref{em.2.2} into the following differential equation
\begin{equation}
   \begin{split}
      \bigg[1 - \frac{\mathcal{C}^2}{u^2}\bigg(1 &
      -\frac{2M}{r}+u^2\bigg)\bigg] u\frac{\ud u}{\ud r} =  \\
      & -\frac{M}{r^2} +
      2\frac{\mathcal{C}^2}{r}\left(1-\frac{2M}{r}+u^2\right).
   \end{split}
   \label{em.df}
\end{equation}
By requiring that both sides of this equation vanish simultaneously at $r_s$,
the following conditions arise
\begin{subequations}
\begin{gather}
   r_s = \frac{1}{2}\frac{M}{u_s^2}, 
   \label{em.?1}\\
   u_s^2 = \frac{\mathcal{C}_s^2}{1+3\,\mathcal{C}_s^2}.
   \label{em.?2}
\end{gather}
\end{subequations}

If we introduce $V$ as the norm of the fluid's three-velocity as measured
by local static observers, given in this case by
\begin{equation}
   V = \left( 1 - \frac{2M}{r} \right)^{-1} \left|\frac{U^r}{U^t}\right|,
   \label{Vzamo}
\end{equation}
from \eqs{em.?1} and \eqref{em.?2} it follows that \mbox{$V_s = \mathcal{C}_s$}, which justifies
calling $r_s$ the sonic radius.

On the other hand, a relationship between the fluid state at infinity and at
the sonic point can be obtained by substituting \eqs{em.?1} and \eqref{em.?2}
into \eq{em.4}. Doing this results in the following cubic equation for $h_s$
\citep[see][Appendix A]{TAH2020}
\begin{equation}
   h_s^3 - (3\gamma -2)h^2_\8 h_s+ 3(\gamma -1)h^2_\8 = 0,
   \label{poly}
\end{equation}
as well as the corresponding equation for the sound speed
\begin{equation}
   \mathcal{C}_s^2 = \frac{1}{3}\left(\frac{h_s^2}{h^2_\8} -1 \right).
\end{equation}

The polynomial in \eq{poly} has three real roots but only one satisfies
$h_s > 1$ and thus has physical meaning.\footnote{As long as $\gamma >
1$ and $h_\8 >1$ the cubic polynomial on the left-hand side of \eq{poly}
is positive for $h_s = 0$ and negative for $h_s = 1$, which implies that
it has three real roots lying in the intervals $(-\infty,0)$, $(0,1)$ and
$(1,\infty)$, respectively. See also~\citet{chaverra2016b,richards2021a} for
alternative ways to characterise the sonic radius using $\mathcal{C}_s^2$.}
This root is given by
\begin{equation}
   h_s = 2\,h\8\,\sqrt{\gamma-\frac{2}{3}}
   \,\sin\left(\Psi+\frac{\pi}{6}\right),
   \label{ea.10}
\end{equation}
where
\begin{equation}
   \Psi = \frac{1}{3}\arccos\left[\frac{3(\gamma-1)}{2\,h_\8}\left(
   \gamma-\frac{2}{3}\right)^{-3/2}\right].
   \label{ea.11}
\end{equation}

Substituting these results back into \eq{em.3}, the mass accretion rate can
be expressed in terms of the asymptotic state of the fluid as
\begin{equation}
   \dot{M}_{\rm M} = 4\pi\,\lambda_{\rm M} M^2
   \frac{\rho_\8}{\mathcal{C}^{3}_\8} ,
   \label{dMM}
\end{equation}
where now the numerical factor $\lambda_{\rm M}$ depends not only on $\gamma$
but also on the asymptotic state of the fluid and is given by
\begin{equation}
   \lambda_{\rm M} = \frac{1}{4}
   \left(\frac{h_s}{h_\8}\right)^{\frac{3\gamma-2}{\gamma-1}} \left(
   \frac{\mathcal{C}_s}{\mathcal{C}_\8}\right)^{\frac{5-3\gamma}{\gamma-1}}.
   \label{lM}
\end{equation}

\begin{figure}
   \centering
   \includegraphics[width=0.98\linewidth]{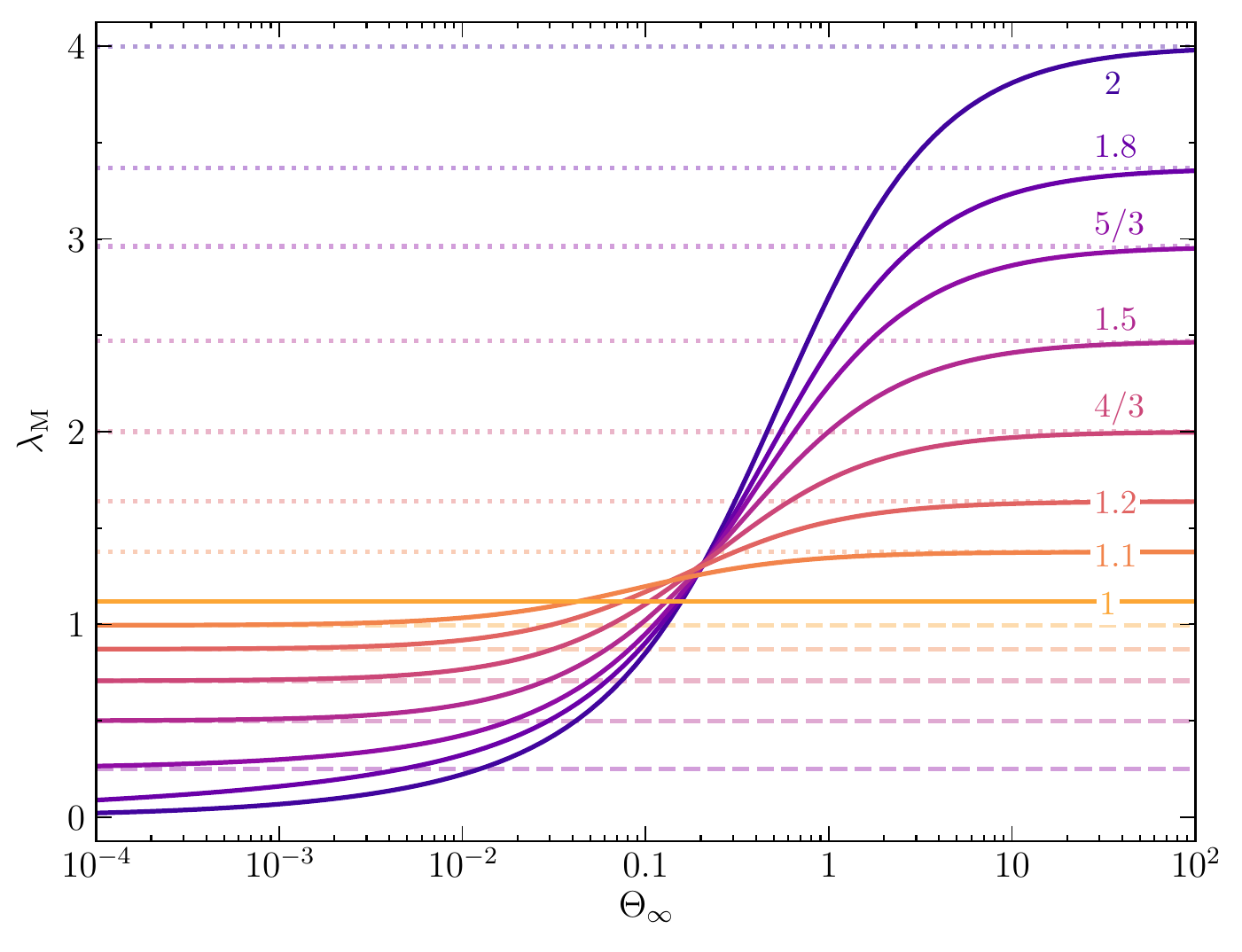}
   \caption{Numerical factor $\lambda_{\rm M}$ in the definition of the
   mass accretion rate of the Michel solution (equation~\ref{lM}) as a
   function of the dimensionless temperature $\Theta\8$ and for different
   values of $\gamma$ as indicated by the labels on top of each curve.
   In the non-relativistic limit $\Theta_\8 \ll1$ and for values of $\gamma
   \le 5/3$, the curves asymptotically approach the values corresponding
   to $\lambda_{\rm B}$ (dashed horizontal lines) of the Bondi solution in
   \eq{bm}. In the ultra-relativistic limit $\Theta_\8 \gg1$, the curves
   asymptotically approach the value given in \eq{lu}.}
   \label{f2}
\end{figure}

In Figure~\ref{f2} we show the dependence of $\lambda_\mathrm{M}$ on
$\Theta_\8$ for several different values of the adiabatic index $\gamma$. From
this figure, it is clear that for $\gamma\le 5/3$ in the non-relativistic
limit $(\Theta_\8 \ll 1)\ \lambda_\mathrm{M} \to \lambda_\mathrm{B}$, as
expected. We stress that the mass accretion rate given in \eq{dMM} is only a
measure for the flux of rest-mass (particle number times the average rest-mass
per particle) onto the central black hole. If interested in computing the
actual growth rate of the black hole's mass, the total energy advected by
each fluid particle should be taken into account by computing the energy
accretion rate \citep[for further details see][]{AST2021}. Since for the
present problem the fluid is assumed to be at rest at infinity, one only needs
to multiply $\dot{M}_{\rm M}$ in \eq{dMM} by $h_\8$ to obtain this rate, i.e.
\begin{equation}
    \dot{\mathcal{E}}_{\rm M} = 4\pi\,\lambda_{\rm M} M^2 \rho_\8
    \left(\frac{1}{\gamma\Theta_\8} + \frac{1}{\gamma-1}\right)^{5/2}
    \gamma\Theta_\8.
    \label{dMM2}
\end{equation}

In contrast to the Bondi solution, where the asymptotic speed of sound
$\mathcal{C}_\8$ is the only characteristic velocity, the Michel solution
naturally features the speed of light as an additional characteristic
velocity. Consequently, the Michel solution can only be rendered scale
invariant with respect to $M$ and $\rho_\infty$. Therefore, in addition to
the adiabatic index $\gamma$, to completely describe a given solution one must
also specify $\mathcal{C}_\8$ or, alternatively, $\Theta_\8$. In what follows
we shall use $\Theta_\8$ as the dynamically relevant parameter describing
the state of the fluid asymptotically far away from the central object.

Examples of the resulting Mach number $\mathcal{M}$ for a \mbox{$\gamma=4/3$}
polytrope and various asymptotic temperatures: \mbox{$\Theta_\8 =
10^{-4},\,0.1,\,10^4$} are also shown in Figure~\ref{fig:mach}. 
Note that, in the relativistic case we have defined 
\begin{equation}
\mathcal{M} = \frac{V\sqrt{1 - \mathcal{C}^2}}{\mathcal{C}\sqrt{1 - V^2}}.
\end{equation}
As we can see from this figure, in all cases the accretion flow transi\-tions
from being subsonic to supersonic before reaching the event horizon 
(indicated by the dashed lines). Also, the non-relativistic limit ($\Theta_\8\ll1$)
coincides with the Bondi solution.
 
\begin{figure}
   \centering
   \includegraphics[width=0.475\textwidth]{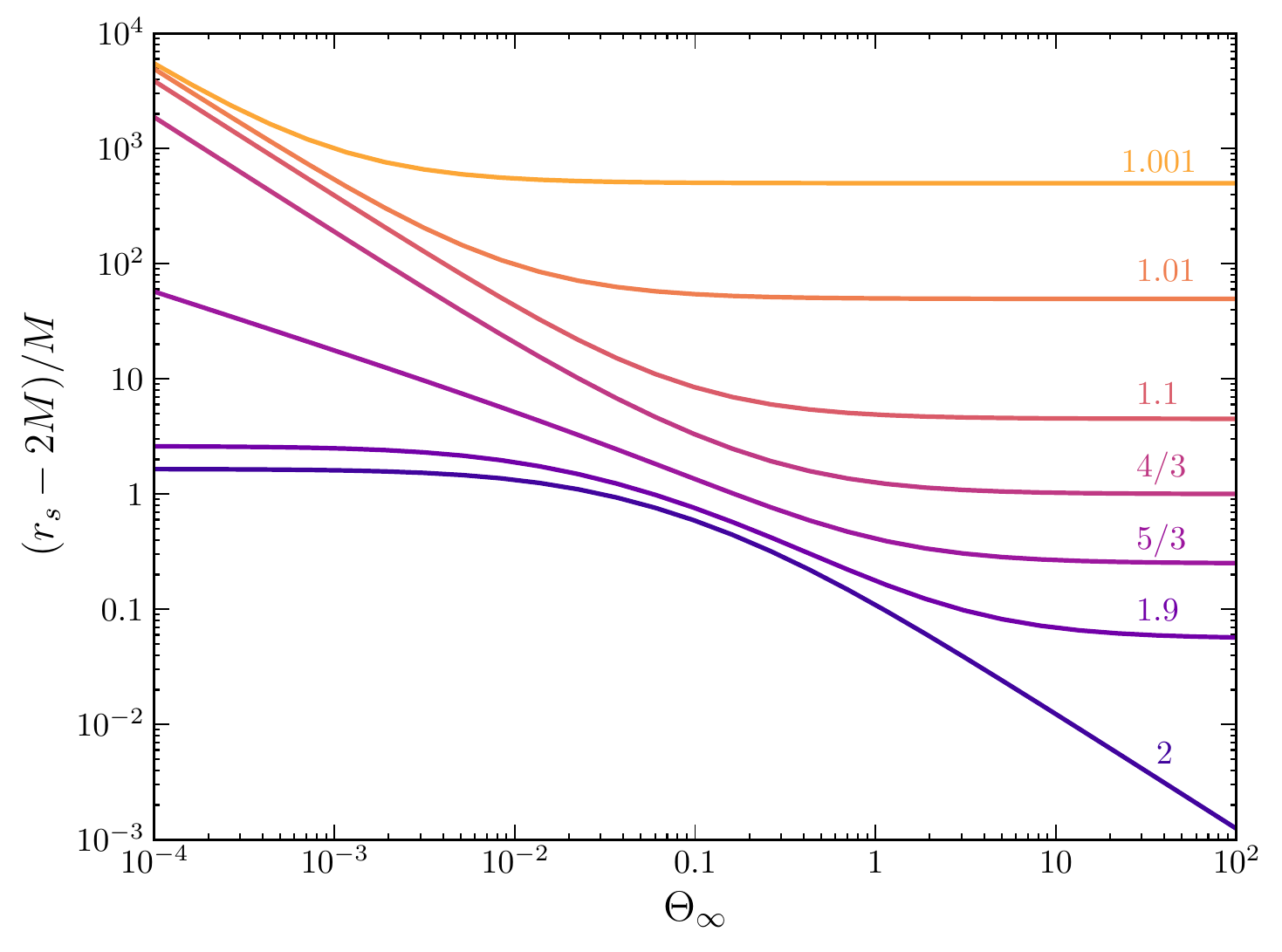}
   \caption{Distance between the sonic radius $r_s$ and the black hole's
   event horizon radius $r_+=2M$ in the Michel solution as a function of the
   dimensionless temperature $\Theta\8$ and for different values of $\gamma$
   as indicated by the labels on top of each curve.}
   \label{f1}
\end{figure}

\begin{figure}
   \centering
   \includegraphics[width=0.475\textwidth]{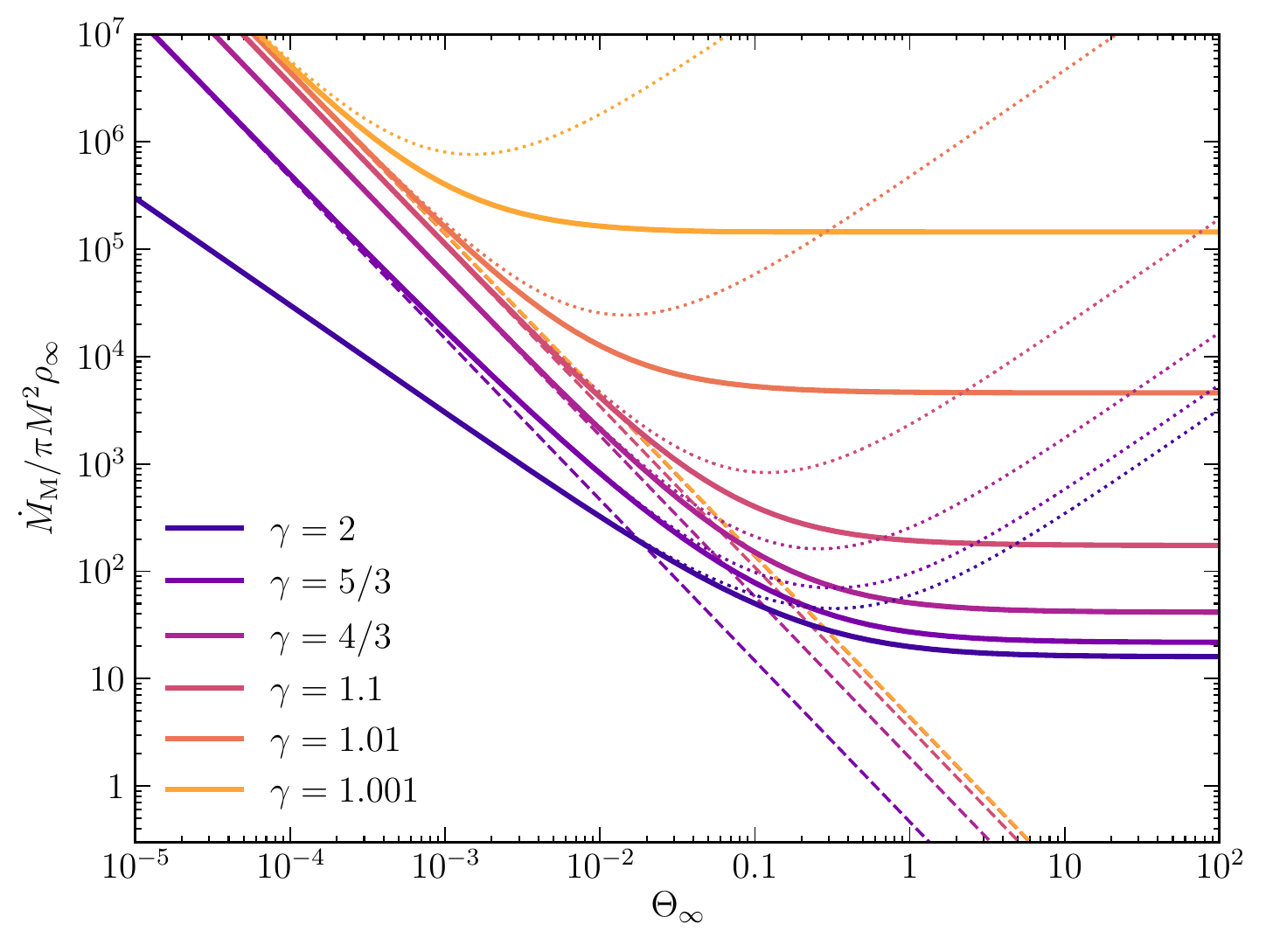}
   \caption{Mass accretion rate of the Michel solution $\dot{M}_\mathrm{M}$ as
   a function of $\Theta_\8$ and for several representative values of $\gamma$.
   The dashed lines represent the corresponding Bondi mass accretion rate
   $\dot{M}_\mathrm{B}$ in cases with $\gamma \le 5/3$. The dotted lines show
   the corresponding energy accretion rate (see equation~\ref{dMM2}). Note
   that once $\Theta_\8 \gtrsim 10^{-2}$, the differences between Bondi's and
   Michel's solutions become of order one and that this difference actually
   diverges as $\Theta_\8\to \infty$.}
   \label{fig:MmMb}
\end{figure}

In Figures~\ref{f1} and \ref{fig:MmMb} we show, respectively, the sonic
radius $r_s$ and the mass accretion rate $\dot{M}_\mathrm{M}$ as functions of
$\Theta_\8$ and for several representative values of $\gamma$. By examining
these figures, and ana\-ly\-sing in detail the result obtained in \eq{dMM},
three interes\-ting limits can be identified (isothermal, non-relativistic and
ultra-relativistic). In what follows, we list the main conclusions that can be
drawn in each case, leaving detailed calculations to Appendix~\ref{app:michel}.

\vspace{8pt}
\noindent {\em (i) Isothermal limit}
\vspace{8pt}

The isothermal limit corresponds to the condition when \mbox{$\gamma \to 1$}.
From Figure~\ref{f1} we note that, within this limit and for all temperatures
$\Theta_\infty$, the sonic radius recedes without limit from the event
horizon. As we prove in Appendix~\ref{app:michel}, the Michel solution (with
a suitable rescaling) converges to the Newtonian Bondi solution with an EoS
as given by \eq{eb.6}. Thus, the isothermal case can be entirely described
within the context of Newtonian physics, even for large temperatures that
would ordinarily be associated with an ultra-relativistic regime.


\vspace{8pt}
\noindent {\em (ii) Non-relativistic limit}
\vspace{8pt}

This limit is described by the condition $\Theta_\infty \ll 1$ which implies
$h_\infty \rightarrow 1$. As expected, and as is already apparent from
Figures~\ref{fig:mach} and \ref{f2}, in this limit the Michel solution
converges to the Bondi one and expressions like the mass accretion
rate (equation~\ref{dMM}) reduce to their non-relativistic counterparts
(equation~\ref{bm}). Nevertheless, this is only true for $\gamma\le 5/3$.
When $\gamma > 5/3$ a qualitative change takes place in Michel's solution. From
Figure~\ref{f1} it is clear that for $\gamma < 5/3$ the value of $r_s$
grows to infinity as $\Theta_\infty^{-1}$ (as in the Bondi solution),
while it converges to a finite distance from the event horizon for $\gamma
> 5/3$. This is indicative that the cases with $\gamma > 5/3$ cannot be
described with Newtonian physics, even in the low temperature limit. As shown
in Appendix~\ref{app:michel}, when $\Theta_\8\ll 1$ and $\gamma>5/3$, one
finds that $r_s$ converges to a finite value and $h_s >1$, while the resulting mass accretion
rate converges to
\begin{equation}
   \dot{M}_{\rm M}\to \pi h_s^{\frac{3\gamma-2}{\gamma-1}}
   \mathcal{C}_s^{\frac{5-3\gamma}{\gamma-1}} M^2\rho_\8
   \,\mathcal{C}^{-\frac{2}{\gamma-1}}_\8.
\end{equation}


\vspace{8pt}
\noindent {\em (iii) Ultra-relativistic limit}
\vspace{8pt}

Finally, we discuss the case where $\Theta_\8\gg1$. From Figure~\ref{f1}
it is clear that $r_s$ converges to a finite value strictly larger than
the event horizon radius for all values of $\gamma$, with the exception
of a stiff EoS $\gamma = 2$, in which case $r_s \to r_+$ as $\Theta_\8 \to
\infty$. Moreover, within this limit \eq{ea.10} reduces to
\begin{equation}
   \frac{h_s}{h_\8} \to \sqrt{3\gamma-2},
\end{equation}
from which one also obtains $\mathcal{C}_s/\mathcal{C}_\8\to 1$ and, hence,
\begin{equation}
   \lambda_{\rm M} \to
   \frac{1}{4} (3\gamma-2)^{\frac{3\gamma-2}{2(\gamma-1)}}.
   \label{lu}
\end{equation}
This limit value grows monotonically from $\mathrm{e}^{3/2}\simeq 1.12$ to
$4$ as $\gamma$ increases from $1$ to $2$ (see Figure~\ref{f2}). From this
result, and as is also clear from Figure~\ref{fig:MmMb}, one sees that the
mass accretion rate becomes independent of $\Theta_\8$, rapidly approaching
the constant value
\begin{equation}
   \dot{M}_\mathrm{M} \to   \pi  M^2 \rho_\8
   (3\gamma-2)^{\frac{3\gamma-2}{2(\gamma-1)}} (\gamma-1)^{-3/2}.  
   \label{dmu}
\end{equation}
In comparison, the Bondi mass accretion rate steadily decreases as
$\Theta^{-3/2}_\8$ as the temperature increases. Therefore, the difference
between $\dot{M}_\mathrm{B}$ and $\dot{M}_\mathrm{M}$ becomes arbitrarily
large when $\Theta_\8\gg1$. Also note that the energy accretion rate is
not monotonic; it decreases for small temperatures but increases for large
ones, eventually growing linearly in $\Theta_\8$ (see equation~\ref{dMM2}
and Figure~\ref{fig:MmMb}). A similar qualitative behaviour has been observed
for the accretion of a Vlasov gas \citep{rioseco2017b}. This is a remarkable
difference between the Bondi and Michel solutions that, to the best of our
knowledge, had not been discussed in the literature before.\footnote{In
the comparison presented in~\citet{malec1999} it is stated that, due to
relativistic effects, the Michel mass accretion is enhanced by, at most,
a factor of 10 as compared to the Bondi value, whereas in our case this
factor is unbounded. Note, however, that the adopted EoS in that work is $P
= K\epsilon^\gamma$, with $\epsilon$ the energy density.} We can track the
reason behind this behaviour by examining \eqs{bm} and \eqref{dMM}. Even
though the factor $\mathcal{C}^{-3}_\8$ appears in both expressions, this
factor behaves drastically differently when $\Theta_\8 \gtrsim 1$. For a
perfect fluid in Newtonian hydrodynamics, one simply has \mbox{$\mathcal{C}^2 =
\gamma\,\Theta$} and, thus, as the temperature increases so does the speed of
sound without limit. Consequently, and as can be seen in Figure~\ref{fig:MmMb},
the Bondi mass accretion rate decreases to small values as the asymptotic gas
temperature increases. On the other hand, in relativistic hydrodynamics one has
$$\mathcal{C}^2 = \frac{\gamma(\gamma-1)\Theta}{\gamma-1 + \gamma\,\Theta},$$
that, in the limit $\Theta_\8\gg 1$, implies that the speed of sound attains
a maximum value given by $\mathcal{C}_\8 \to \sqrt{\gamma -1}$. Therefore,
when \mbox{$\Theta_\8 \gg 1$}, $\dot{M}_\mathrm{M}$ becomes independent
of $\Theta_\8$.

\subsection{Michel solution with relativistic EoS}
\label{subsec:releos}

In the previous subsection we revisited spherical accretion of a fluid
that follows an ideal gas EoS and that is restricted to obey a polytropic
relation. As mentioned before, assuming a monoatomic gas, this restriction is
only valid in the non-relativistic limit ($\Theta_\8 \ll 1$) with $\gamma =
5/3$ or in the ultra-relativistic one ($\Theta_\8 \gg 1$) with $\gamma =
4/3$. Nevertheless, as it was shown by~\citet{taub1948}, in the relativistic
case ($\Theta_\8 \sim 1$) the polytropic restriction is not physical and
has to be dropped.

In this subsection we extend the Michel solution to the case of a gas
obeying an appropriate EoS for the relativistic regime. As derived from
relativistic kinetic theory, the EoS of an ideal, monoatomic gas can be
written as~\citep{juttner1911,synge1957,falle1996}
\begin{equation}
   h = \frac{ K_3(1/\Theta)}{ K_2(1/\Theta)},
   \label{e.2.3.h}
\end{equation}
where, as before $\Theta = P/\rho$, and $K_n$ is the $n$th-order modified
Bessel function of the second kind.\footnote{We adopt the definition of the
modified Bessel function as presented in \url{https://dlmf.nist.gov/10.25}.}

By additionally imposing the adiabatic condition (i.e.~isentropic flow),
one obtains the following relation between $\rho$ and $\Theta$ \citep[see
e.g., Appendix B of][]{chavez2020}
\begin{subequations}
\begin{gather}
   \frac{\rho}{\rho_\8} = \frac{f(\Theta)}{f(\Theta_\8)}, \label{e.2.3.rho}\\
   f(\Theta) = \Theta\,K_2(1/\Theta)\,
   \exp\left[\frac{1}{\Theta}\frac{ K_1(1/\Theta)}{ K_2(1/\Theta)} \right].
\end{gather}
\end{subequations}
Meanwhile, the speed of sound in this case is given by 
\begin{equation}
   \mathcal{C}^2 = \frac{\bar\gamma\,\Theta}{h},
   \label{e.2.3.sound}
\end{equation}
where $\bar\gamma$ is the effective adiabatic index, defined as
\begin{equation}
   \bar\gamma := \frac{\rho}{P}\frac{\partial P}{\partial \rho} =
   \frac{h}{\Theta} \mathcal{C}^2.
\end{equation}

In contrast to the polytropic gas treatment discussed before, $\bar\gamma$
is not a constant but rather a function of the temperature that can be
calculated explicitly as
\begin{equation}
   \bar\gamma = \frac{h'}{h' + \Theta^2},
\end{equation}
where the prime refers to derivatives with respect to the argument of
the modified Bessel functions, i.e.~$h' = \ud [ K_3(x)/ K_2(x) ] / \ud
x$. With this definition of $\bar\gamma$ it follows that, as expected,
for non-relativistic temperatures, $\bar\gamma \to 5/3$ while, in the
ultra-relativistic limit, $\bar\gamma \to 4/3$.

In order to derive the appropriate governing equations in this case, we
first notice that \eq{poly} should be replaced with
\begin{equation}
   h_s^2 = h^2_\8(1 + 3\,\mathcal{C}_s^2) = 
   h^2_\8\left[1 + \frac{3\Theta_s}{h_s} \left( \frac{h_s'}{h_s' +
   \Theta_s^2}\right)\right],
   \label{poly2}
\end{equation}
which, in contrast to \eq{poly}, does not allow for an analytic
solution. Nevertheless, it can be easily solved numerically using any standard
root finding algorithm.

The corresponding mass accretion rate is obtained by evaluating \eq{em.3}
at the sonic point, i.e.
\begin{equation}
   \dot{M} = 4\pi\,r_s^2\,\rho_s\,u_s ,
\end{equation}
and, by applying the conditions given by \eqs{em.?1} and \eqref{em.?2} that,
together with \eq{e.2.3.rho}, result in
\begin{equation}
   \dot{M} =  \pi M^2 \rho_\8
   \frac{(1+3\,\mathcal{C}_s^2)}{\mathcal{C}_s^3}^{3/2}
   \frac{f(\Theta_s)}{f(\Theta_\8)}.
   \label{dmrel}
\end{equation}
In practice, to calculate the resulting mass accretion rate for a given
asymptotic state $(\rho_\8,\Theta_\8)$, we numerically solve \eq{poly2}
to obtain $\Theta_s$, from which we can compute $h_s$ and $\mathcal{C}_s$
via \eqs{e.2.3.h} and \eqref{e.2.3.sound}, respectively, and then substitute
these values into \eq{dmrel}.

\begin{figure}
   \centering
   \includegraphics[width=0.475\textwidth]{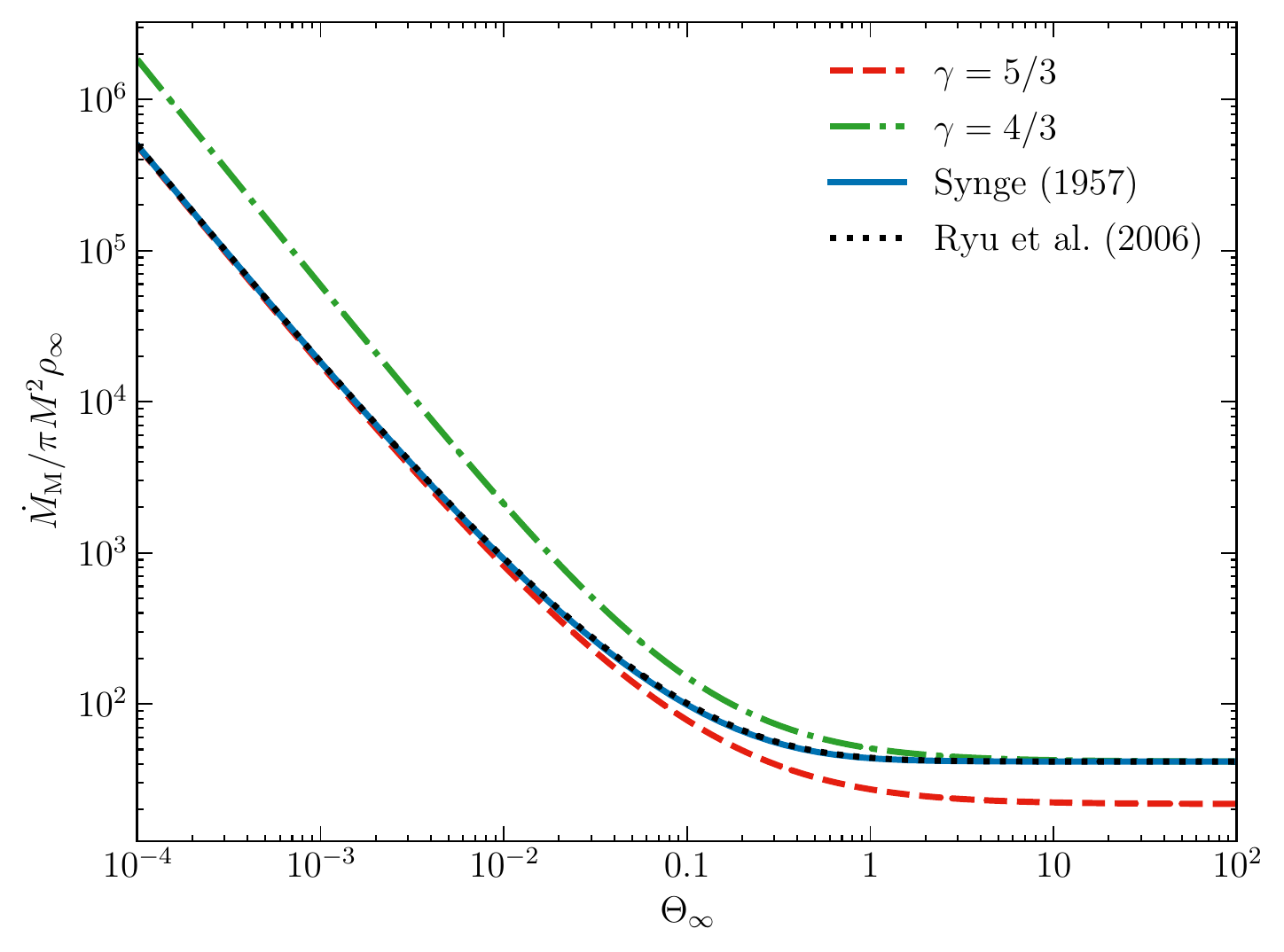}
   \caption{Mass accretion rate in Michel's model for a gas described by
   a relativistic EoS \citep{synge1957}. The resulting rate converges to
   a $\gamma = 5/3$ polytrope when $\Theta_\8 \ll 1$ while it behaves as a
   $\gamma = 4/3$ polytrope for $\Theta_\8 \gg 1$. Also shown is the result
   of using the approximation to the relativistic EoS by~\citet{ryu2006}.}
   \label{f_synge}
\end{figure}

In Figure~\ref{f_synge} we show the resulting mass accretion rate as a
function of $\Theta_\8$ for the relativistic EoS and compare it with the
corresponding values for $\gamma=5/3,\,4/3$ polytropes. From this figure
we can see that the result obtained with the relativistic EoS provides a
smooth transition between the polytropic approximations as the temperature
transitions from non-relativistic values to the ultra-relativistic regime.   
We also show the approximation to the relativistic EoS proposed
by~\cite{ryu2006}, where,
\begin{equation}
   h = 2\frac{6\Theta^2 + 4\Theta + 1}{3\Theta + 2},
   \label{eq:ryuh}
\end{equation}
and which provides an accurate estimate to the mass accretion rate to
within 2$\%$. This comparison is relevant for this work, given that, for
some of the numerical simulations presented in Section~\ref{sec:polytropic},
we have adopted this proxy for the implementation of the relativistic EoS.

\subsection{Ultra-relativistic, stiff fluid in Kerr spacetime}
\label{subsec:stiff}

The analytic solutions revisited so far consider a non-rotating black hole
as the central accretor. The inclusion of the black hole's spin breaks
the spherical symmetry of the problem, resulting in a new scenario for
which it is not clear whether it admits a closed, analytic solution in
general.\footnote{Both \cite{shapiro1974} and \citet{zanotti2005} have
proposed a Michel-like solution for spherical accretion onto a rotating Kerr
black hole that is built on the assumption that the polar angular velocity
vanishes everywhere. However, as we show in Section~\ref{sec:polytropic},
this condition is not satisfied for a general perfect fluid.} As mentioned
in the Introduction, a notable exception is the solution derived by
\citet*{petrich1988} (PST henceforth), that we shall now briefly review. In
that work, the authors found a full analytic solution for accretion onto
a Kerr black hole which is, however, restricted to the special case of an
ultra-relativistic stiff fluid. Within this approximation, the fluid rest-mass
energy is neglected as compared to its internal energy, while the stiff
condition means that a $\gamma=2$ polytrope is being considered. Under these
conditions, the thermodynamic variables of the fluid are simply related as
\begin{equation} 
   P=K\rho^2, \qquad h = 2\,K\,\rho. 
   \label{es.1}
\end{equation}
Moreover, the spacetime metric is considered as fixed and corresponding to
a Kerr black hole of mass $M$ and spin parameter $a$, in other words, the
accreting gas is assumed to be a test fluid with a negligible self-gravity
contribution. With the further assumptions of steady-state and irrotational
flow, the fluid is described as the gradient of a scalar potential $\Phi$
such that
\begin{equation}
   h\,U_\mu = \Phi_{,\mu},
   \label{es.2}
\end{equation}
and, by imposing the normalization condition of the four-velocity,
\begin{equation}
   h = \sqrt{-\Phi_{,\mu}\Phi^{,\mu}}.
   \label{es.3}
\end{equation}

By substituting \eq{es.2} into \eq{em.1}, it follows that $\Phi$ satisfies
the linear wave equation
\begin{equation}
   \Phi_{,\mu}^{\hspace{7pt};\mu} = \frac{1}{\sqrt{-g}}\left(
   \sqrt{-g}\,g^{\mu\nu} \Phi_{,\mu} \right)_{,\nu} = 0,
   \label{es.4}
\end{equation}
where $g^{\mu\nu}$ and $\sqrt{-g}$ are, respectively, the inverse and the
determinant of the Kerr metric. In what follows we shall adopt Kerr-type
coordinates $(t,\,r,\,\theta,\,\phi)$ in which the line element assumes
the form
\begin{equation}
\begin{split}
   \ud s^2 =\ & -\left(1-\frac{2Mr}{\varrho^2}\right)\ud t^2 +
   \left(1+\frac{2Mr}{\varrho^2}\right)\ud r^2 \\
   & + \frac{4Mr}{\varrho^2}\ud t\, \ud r -
   \frac{4\,aMr}{\varrho^2}\sin^2\theta\,\ud t\, \ud \phi \\
   & -2\,a\left(1+\frac{2Mr}{\varrho^2}\right)\sin^2\theta\,\ud r\, \ud \phi \\
   & + \varrho^2\ud\theta^2 
   + \frac{\Sigma\,\sin^2\theta}{\varrho^2}\ud\phi^2,
   \end{split}
   \label{es.5}
\end{equation}
with the functions\footnote{We use the same notation as \citet{AST2021} and
warn the reader that the symbol $\varrho$ refers to the metric coefficient
defined in \eq{es.6} which should be distinguished from the similar-looking
symbol $\rho$ which denotes the rest-mass density.}
\begin{subequations}
   \begin{gather}
      \varrho^2 = r^2 + a^2\cos^2\theta, 
      \label{es.6}\\
      \Sigma = \left(r^2 + a^2\right)^2 - a^2\Delta\sin^2\theta, 
      \label{es.7}\\
      \Delta = r^2 - 2Mr + a^2. 
      \label{es.8}
   \end{gather}
\end{subequations}

By requiring that the fluid is uniform and at rest asymptotically far away
from the central object, the solution is given by \citet{AST2021}
\begin{equation}
   \Phi = h_\8\left[ -t + 2M\ln\left(\frac{r - r_-}{r_+ - r_-} \right) \right],
   \label{es.9}
\end{equation}
where $r_\pm = M \pm \sqrt{M^2-a^2}$ are the roots of the equation $\Delta =
0$, with $r_+$ corresponding to the event horizon and $r_-$ to the Cauchy
horizon of the Kerr black hole. It is clear that $\Phi$ is regular everywhere
outside the Cauchy horizon $r>r_-$.

Substituting the velocity potential in \eq{es.9} into \eq{es.2}, leads to 
\begin{subequations}
\begin{align}
   & \frac{h}{h_\8}U^t = 1 + \frac{2M r}{\varrho^2}\left(\frac{r + r_+}{r
   - r_-}\right), 
   \label{es.10}\\
   & \frac{h}{h_\8}U^r = -\frac{2Mr_+}{\varrho^2}, 
   \label{es.11}\\
   & \frac{h}{h_\8}U^\theta = 0, 
   \label{es.12}\\
   & \frac{h}{h_\8}U^\varphi = \frac{2\,aMr}{\varrho^2(r-r_-)}, 
   \label{es.13}
\end{align}
\end{subequations}
while, by combining \eqs{es.1} and \eqref{es.3}, one obtains
\begin{equation}
   \frac{\rho}{\rho_\8} = \frac{h}{h_\8}  = \sqrt{ 1 +
   \frac{2M}{\varrho^2}\frac{r(r + r_+) + 2M r_+}{r - r_-} }.
   \label{es.14}
\end{equation}
 
Note that, although the fluid's four-velocity has a non-vanishing
azimuthal component when $a\neq 0$, its angular momentum is zero
since $U_\mu\xi^\mu_{(\phi)} = U_\phi = 0$, where $ \xi^\mu_{(\phi)} =
\delta^\mu_3$ is the Killing vector field associated with the axisymmetry
of Kerr spacetime. Also note that, both the four-velocity and the fluid
density, are well-defined for all $r > r_-$ (including at the event horizon)
but diverge as one approaches the Cauchy (inner) horizon $r\to r_-$.

In the non-rotating case \eq{es.14} reduces to
\begin{equation}
   \rho = \rho_\8\sqrt{ 1 + \frac{2M}{r} + \left( \frac{2M}{r} \right)^2 +
   \left( \frac{2M}{r} \right)^3},
   \label{es.15}
\end{equation}
which agrees with the findings in Section~4.2 of \cite{chaverra2015},
with a compression rate of $\rho(r_+)/\rho_\8 = 2$ at the horizon. In
the rotating case, this compression rate can be considerably higher, with
$\rho(r_+)/\rho_\8\to \infty$ in the maximally rotating limit $|a|\to M$.

The resulting mass accretion rate for the potential flow described by \eq{es.9}
is given by
\begin{equation} 
   \dot{M}_{\rm PST} =  8\pi M r_+ \rho_\8 = 4\pi(r_+^2+a^2)\rho_\8.
   \label{es.16}
\end{equation}
Interestingly, from \eq{es.16} we see that, in this special case of
an ultra-relativistic stiff fluid, the resulting mass accretion rate is
proportional to the event horizon area $A=4\pi(r_+^2+a^2)$ \citep{carroll2003},
and, consequently, for fixed $M$ and $\rho_\8$, $\dot{M}_{\rm PST}$
decreases as $|a|$ increases, having the finite limit $\dot{M}_{\rm PST}
=8\pi M^2\rho_\8$ when $|a|\to M$.  We also note that, for a non-rotating
black hole, $\dot{M}_{\rm PST} =16\pi M^2\rho_\8$, which coincides exactly
with the result given in \eq{dmu} when $\gamma=2$.

One inconvenience of assuming an ultra-relativistic stiff EoS, is that the
speed of sound equals the speed of light, leading to a model with a limited
applicability in astrophysics. Nevertheless, it represents a fully hydrodynamic
exact solution that is very useful as a benchmark test for the validation of
general relativistic hydrodynamic numerical codes in a fixed Kerr spacetime.
In the next section we relax this restriction on the EoS.

\setcounter{equation}{0}
\section{Perfect fluid in Kerr spacetime}
\label{sec:polytropic}

In the previous sections we reviewed, along with the Bondi and Michel models,
the analytic PST solution. This is the only exact solution that considers
a rotating black hole as central accretor. This solution corresponds to an
upper limit in both the temperature of the gas $(\Theta_\infty \gg 1)$ and
in the adiabatic index ($\gamma = 2$). Unfortunately, for a more general EoS,
or even just a different value of $\gamma$, it is apparently not possible to
find a closed analytic solution. Therefore, we explore the spherical accretion
of a perfect fluid with a more general EoS onto a rotating Kerr black hole by
means of general relativistic hydrodynamic numerical
simulations.\footnote{Recall that by `spherical accretion onto a rotating
black hole' we mean a solution which is asymptotically spherically symmetric.}
Specifically, we shall focus on the dependence of the resulting accretion
flow on the spin parameter $a$, the asymptotic gas temperature $\Theta_\8$,
and the fluid EoS. We also compare the results with the analytic solutions
presented in Section~\ref{sec:analytic}.

\subsection{Numerical setup and code description}
\label{subsec:numerical}

We perform a total of 311 numerical simulations using the
open source code \textsc{aztekas}.\footnote{The code can be
downloaded from \url{https://github.com/aztekas-code/aztekas-main}.
See~\citet{AMO2018,TA2019,ATH2019,TAH2020}, for further details
regarding the characteristics, test suite and discretization method of
\textsc{aztekas}.} This code solves the general relativistic hydrodynamic
equations, written in a conservative form using a variation of the ``3+1
Valencia formulation''~\citep{banyuls1997} for time independent, fixed
metrics~\citep{delzanna2007}. The spatial integration is carried out using a
grid-based, finite volume scheme coupled with a high resolution shock capturing
method for the flux calculation, and a monotonically centred second order
spatial reconstructor. The time integration is performed using a second order
total variation diminishing Runge-Kutta method~\citep{shu1988}. The evolution
of the equations is performed on a Kerr background metric, using the same
horizon penetrating Kerr-type coordinates as in~Section~\ref{subsec:stiff}.

The set of primitive variables used in the code consists of the rest-mass
density $\rho$, pressure $P$, and the three-velocity vector $v_i$ as measured
by Local Eulerian Observers associated with the chosen coordinate system. Both
$\rho$ and $P$ are thermodynamic quantities measured at the co-moving reference
frame, and the vector $v_i$ is computed as $v_i = \gamma_{ij} v^j$ where
\begin{equation}
    v^i = \frac{U^i}{\alpha U^t} + \frac{\beta^i}{\alpha}, \qquad i = r, \theta, \phi
\end{equation}
with $\alpha$, $\beta^i$ and $\gamma_{ij}$ the lapse, shift vector and
three-metric of the 3+1 formalism~\citep{alcubierre2008}, respectively.

\subsection{Initial and boundary conditions}
\label{subsec:initial_bound}

For all the simulations we adopt a spherical two-dimensional
axisymmetric 2.5D\footnote{The 2.5D scheme consists in evolving
the full 3D system of equations, but imposing the condition that
the fields are independent of $\phi$, such that it is sufficient to
consider a two-dimensional grid. The code is not precisely 2D because
the azimuthal component $v^\phi$ of the three-velocity is allowed to evolve
instead of being set to zero.} domain with coordinates $(r,\theta) \in
[\mathcal{R}_{\mathrm{in}},\mathcal{R}_{\mathrm{out}}]\times [0,\pi/2]$,
where $\mathcal{R}_{\mathrm{in}}$ and $\mathcal{R}_{\mathrm{out}}$ are the
inner and outer radial boundaries, respectively. We use a uniform polar
grid and an exponential radial grid~\citep[see][for details]{ATH2019} and
fix the numerical resolution to 128$\times$64 grid cells, unless otherwise
stated. Reflective boundaries are set at $\theta=0$ and $\theta = \pi/2$. The
inner radial boundary, at which we impose a free-outflow condition, is
placed within the event horizon ($\mathcal{R}_{\mathrm{in}} < r_+$). On
the other hand, the outer radial boundary is set with the corresponding
Michel solution. With this external boundary condition, the domain size
must be sufficiently large as to avoid introducing numerical artefacts in
the resulting steady-state solution. By performing a quantitative study
varying $\mathcal{R}_{\mathrm{out}}$, we find that we can be confident of
the independence on the domain size by taking $\mathcal{R}_{\mathrm{out}}
= 10 \, r_\mathrm{B}$ in the non-relativistic regime $\left( \Theta_\8
\lesssim 10^{-2} \right)$, and $\mathcal{R}_{\mathrm{out}} = 40 \,
r_\mathrm{s}$ in the relativistic one $\left( \Theta_\8 \gtrsim 10^{-2}
\right)$, where $r_\mathrm{B}$ and $r_\mathrm{s}$ are the Bondi and
sonic radii, respectively. In other words, for a given $\Theta_\8$, we set
$\mathcal{R}_{\mathrm{out}} = 10\, \max( r_\mathrm{B},\ 4\, r_\mathrm{s})$. In
what regards the initial conditions, we start our simulations with a static
$(v_i = 0)$ and uniform $\left( \rho = \rho(\mathcal{R}_{\mathrm{out}}),
P = P(\mathcal{R}_{\mathrm{out}}) \right)$ gas distribution.

The mass accretion rate evolves as a function of time with periodic and
exponentially damped oscillations (in agreement with the results from
\citeauthor{ATH2019}, \citeyear{ATH2019}). The numerical simulations are left
to run until the time variation of the resulting mass accretion rate drops
below 1 part in $10^4$, a criterion that we take as signalling the onset of
the steady-state condition. We compute the mass accretion rate according to
\begin{equation}
    \dot{M} = 4\pi \int_0^{\pi/2} \rho \, \Gamma  \left( v^r -
    \frac{\beta^r}{\alpha} \right) \sqrt{-g} \, \mathrm{d}\theta,
\end{equation}
where $\Gamma = 1/\sqrt{1 - \gamma_{ij}v^iv^j}$ is the Lorentz factor. 

\subsection{Code validation}
\label{subsec:code}

In order to validate our numerical results, we exploit the known analytic
solutions discussed in Section~\ref{sec:analytic} and use them as benchmark
in our test runs.

\begin{figure}
   \centering
   \includegraphics[width=0.475\textwidth]{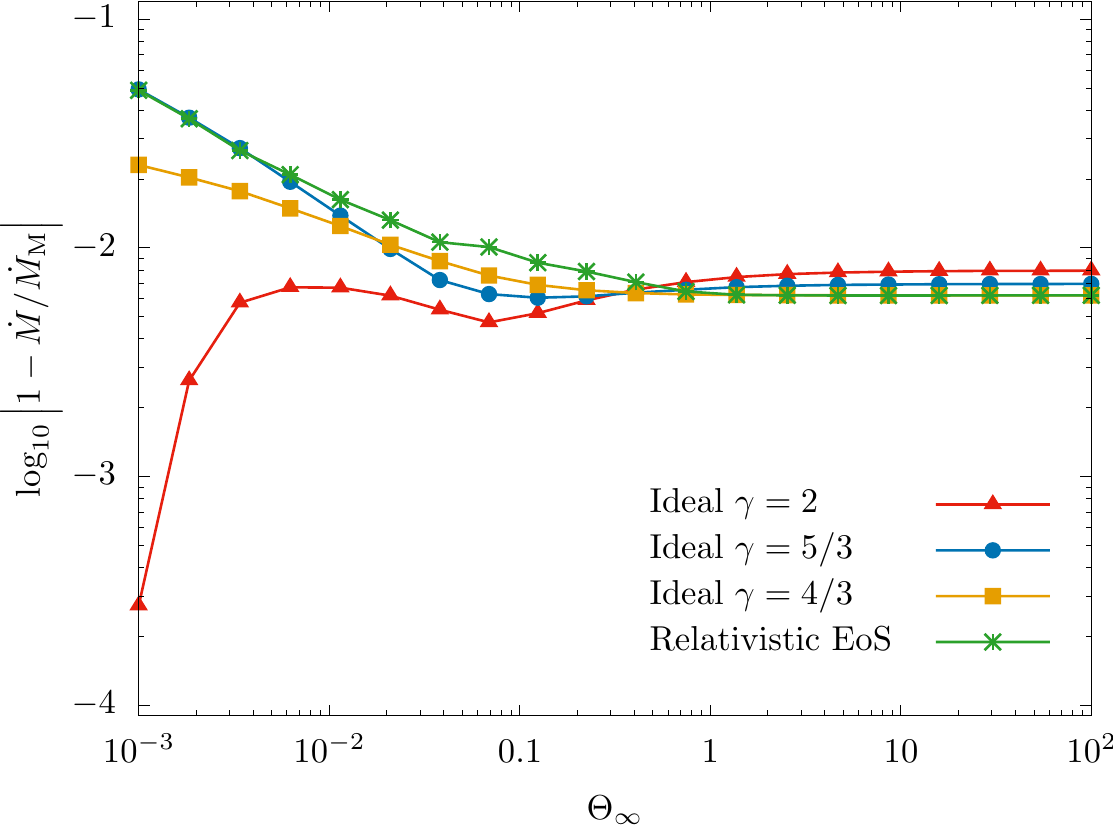}
   \caption{ Relative error in the mass accretion rate between the numerical
   results $(\dot{M})$ and the Michel analytic solution $(\dot{M}_\mathrm{M})$,
   as a function of the asymptotic temperature $\Theta_\8$. We plot the
   results for $\gamma = 4/3, 5/3, 2$ and the relativistic EoS.}
   \label{fig:mdota0}
\end{figure}

Considering first a non-rotating black hole, in Figure~\ref{fig:mdota0} we
show the relative error in the steady-state mass accretion rate between the
Michel analytical solution $(\dot{M}_\mathrm{M})$ and the numerical results
as a function of the asymptotic temperature $\Theta_\8$. We show the results
for $\gamma = 4/3, 5/3, 2$ as well as the fit to the relativistic EoS given
by~\citet{ryu2006}. For simplicity, in what follows we shall refer to this
fit as the ``relativistic EoS". In all cases the numerical error is less than
5\% in the non-relativistic regime ($\Theta_\8 \ll 1$) and less than 1\%
in the relativistic regime ($\Theta_\8 \gg 1$), which is consistent with
the numerical resolution being used.

We also perform additional numerical tests to validate the implementation
of a non-zero spin parameter in our setup. In order to approximate the
ultra-relativistic stiff EoS and to compare with the PST analytic solution,
we perform simulations using an adiabatic index $\gamma = 2$ and an asymptotic
temperature $\Theta_\8 = 10^2$, for different values of $a$. The result of
this comparison is shown in Figures~\ref{fig:mdota99} and~\ref{fig:mdot_vs_a},
from where we find an excellent agreement between both solutions, with a
relative error of less than 1\%. We explain these two figures in further
detail in the next subsection.

\subsection{Results}

In order to quantify the spherical accretion flow onto a rotating Kerr black
hole and analyse its dependence on the black hole's spin parameter $a$, we
perform a series of simulations varying both $a$ and $\Theta_\8$, both for
a polytrope with $\gamma = 4/3, 5/3, 2$ as well as for the relativistic EoS.

For the spin parameter, we take a uniformly distributed set of values between
$a = 0$ (non-rotating black hole) and $a/M = 0.99$. On the other hand, for the
temperatures we choose a list of representative values between $\Theta_\8 =
10^{-3}$ and $10^{2}$, in order to study the behaviour of the solution in
the transition from the non-relativistic regime to the ultra-relativistic one.

\subsubsection{Temperature dependence: rotating black hole case}

We explore the variation in the mass accretion rate for the rotating black
hole case $a > 0$, as compared with the non-rotating case. We find that the
larger difference is obtained for a maximally rotating black hole, as is to
be expected considering the analytic PST solution~(see equation \ref{es.16}).

\begin{figure}
   \centering
   \includegraphics[width=0.475\textwidth]{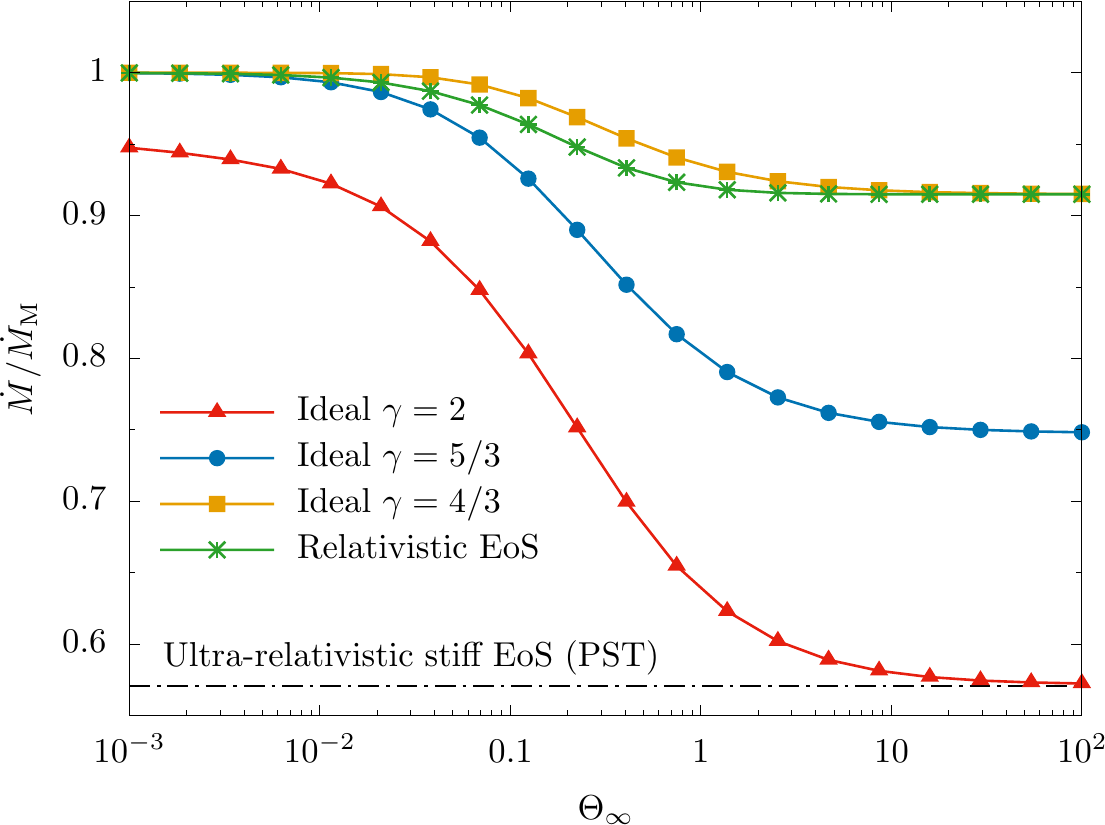}
   \caption{Mass accretion rate as a function of $\Theta_\8$, for a rotating
   black hole with $a/M=0.99$. The first three lines show the results for
   an ideal gas EoS with different values of $\gamma$, and the last line
   represents the relativistic EoS.  The black dashed line represents
   the ultra-relativistic stiff EoS lower limit. The mass accretion rate
   is normalised using the corresponding value in the non-rotating case
   ($\dot{M}_{\rm{M}}$).}
   \label{fig:mdota99}
\end{figure}

In Figure~\ref{fig:mdota99} we show the steady-state mass accretion rate as
a function of the asymptotic temperature (for $\gamma=4/3, 5/3, 2$ and the
relativistic EoS) for the case of a rotating black hole with a spin parameter
$a/M = 0.99$. The mass accretion rate is normalised by the corresponding
Michel value ($a = 0$). As can be seen from this figure, all simulations are
bounded between the non-rotating black hole value $(\dot{M}/\dot{M}_{\rm{M}}
= 1)$ and the ultra-relativistic stiff EoS case $(\dot{M}/\dot{M}_{\rm{M}}
\simeq 0.57)$. In the non-relativistic regime $(\Theta_\infty \ll 1)$, the
mass accretion rate for all $\gamma$ values converges to the corresponding
Michel solution, although this convergence appears to be much slower in
the case $\gamma=2$. Thus, we conclude that in this regime the effects
of the spin on $\dot{M}$ are negligible for $\gamma \leq 5/3$. In the
ultra-relativistic regime $(\Theta_\infty \gg 1)$, $\dot{M}$ decreases by a
factor of $\sim 10$, $25$, and $43\%$ for the solutions with $\gamma = 4/3$,
$5/3$, and $2$, respectively. Note how the solution for $\gamma = 2$ in the
$\Theta_\8 \gg 1$ limit matches the ultra-relativistic stiff analytical value.

\subsubsection{Spin dependence}
\label{S3.4.2}

\begin{figure}
   \centering
   \includegraphics[width=0.475\textwidth]{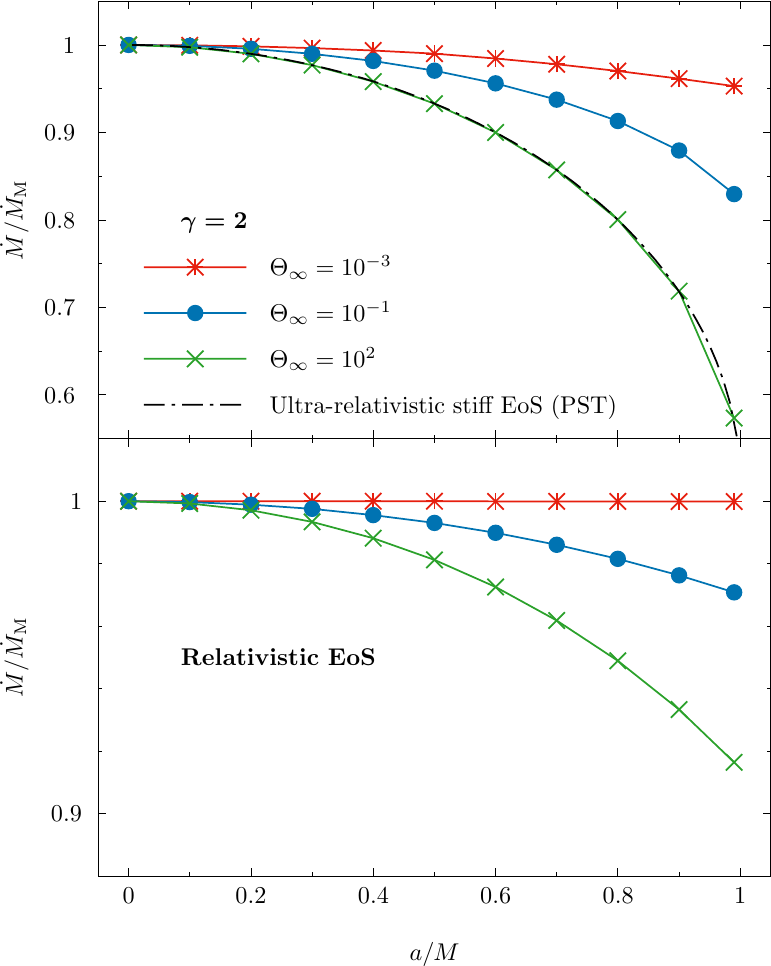}
   \caption{Mass accretion rate as a function of the spin parameter for
   the $\gamma = 2$ (top panel) and the relativistic EoS (bottom panel),
   and different values of the asymptotic temperature $\Theta_\8$. The
   mass accretion rate is normalised by its value in the non-rotating case
   $\dot{M}_{\rm{M}}$. The black dashed line in the top panel represents
   the solution obtained with the ultra-relativistic stiff EoS (PST) model.}
   \label{fig:mdot_vs_a}
\end{figure}

\begin{figure*}
   \centering
   \includegraphics[width=0.475\textwidth]{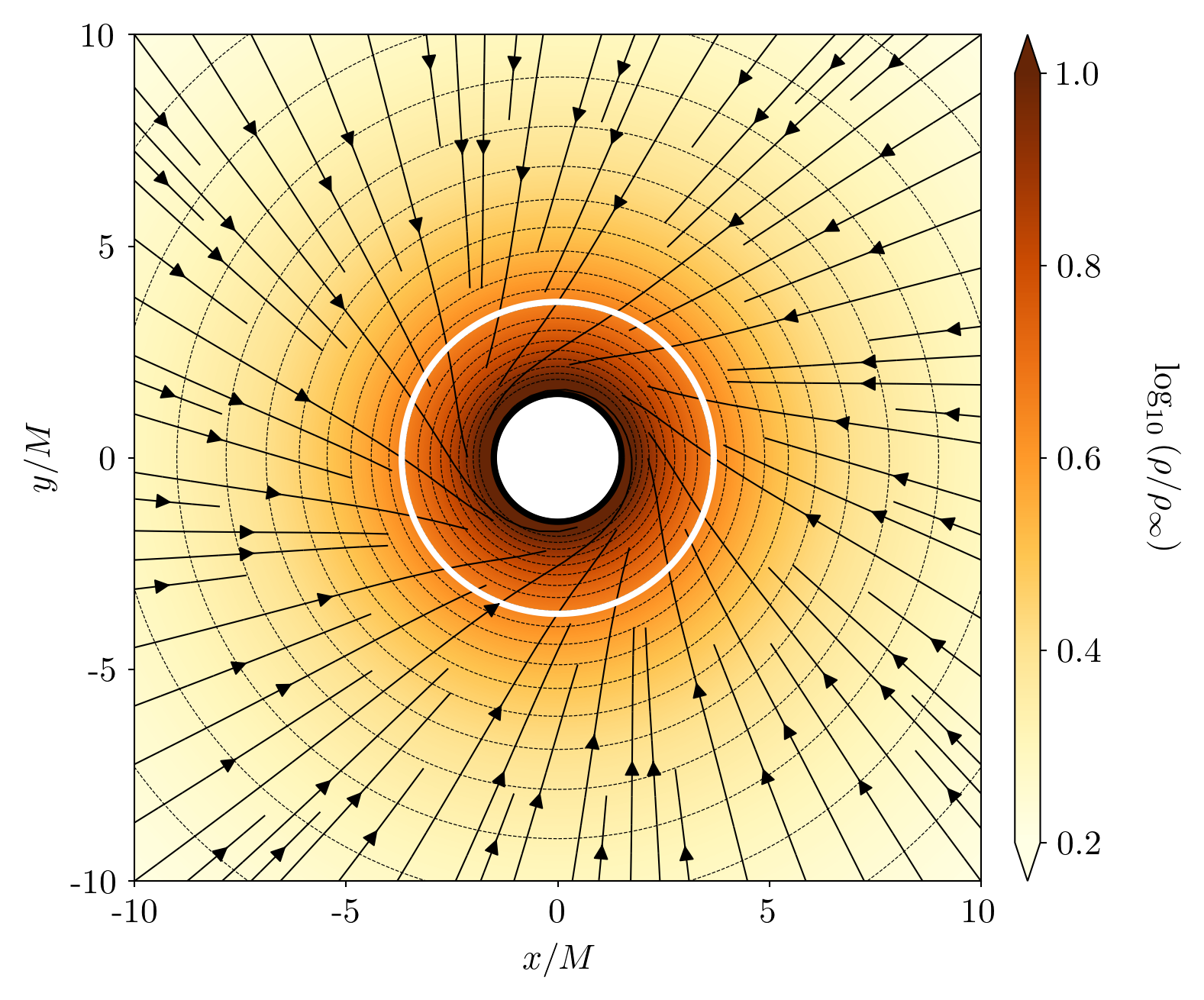}
   \includegraphics[width=0.475\textwidth]{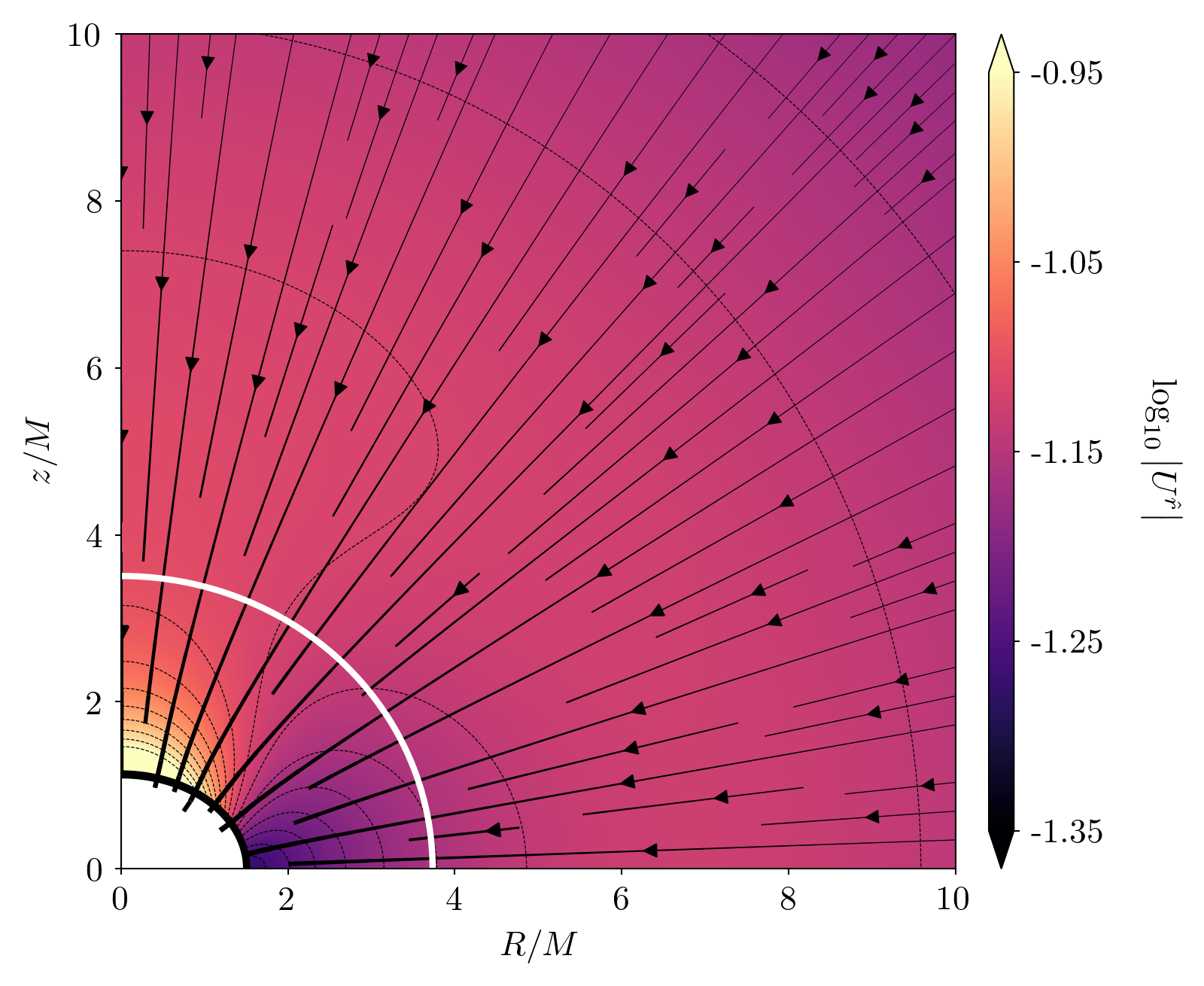}
   \includegraphics[width=0.475\textwidth]{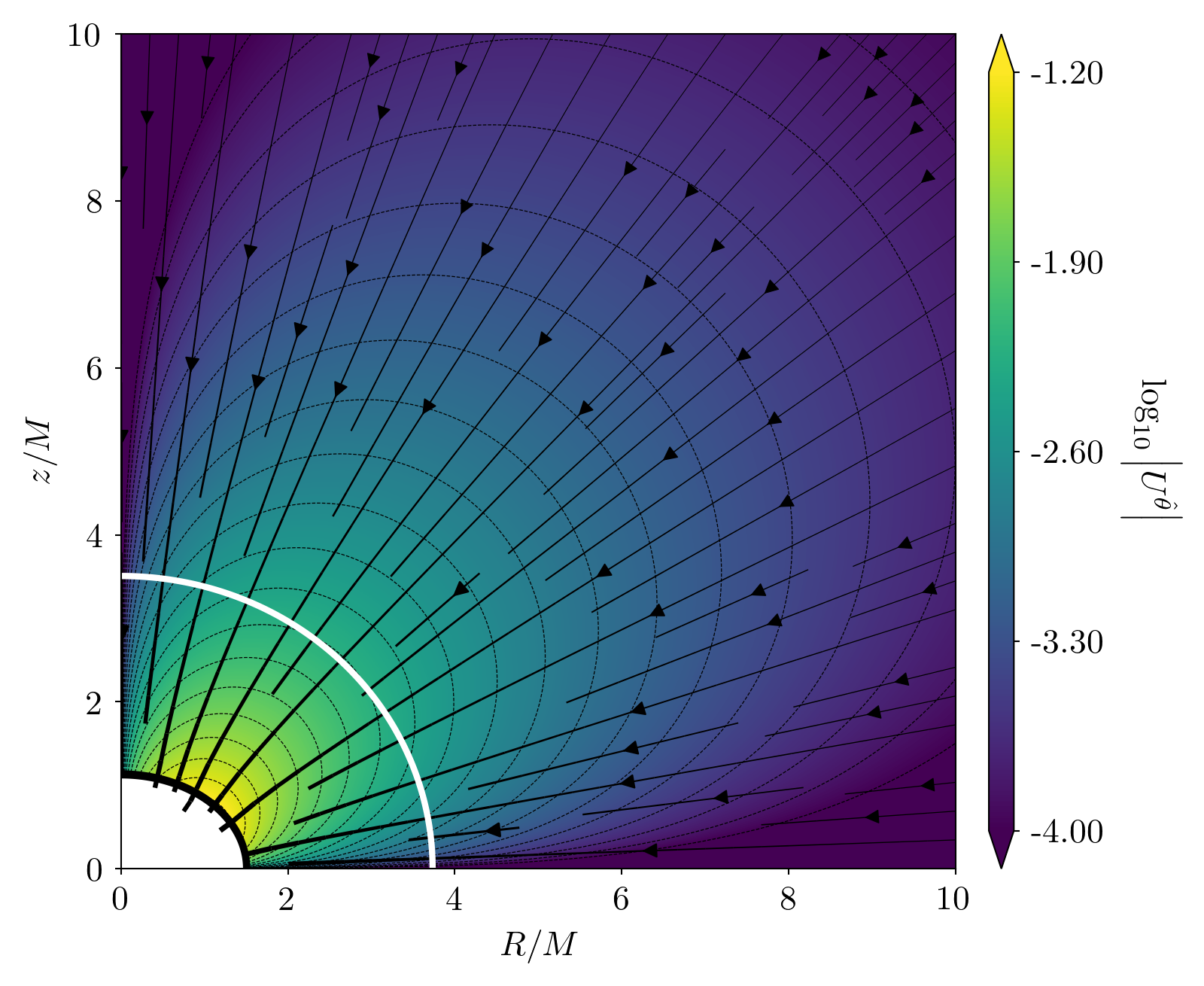}
   \includegraphics[width=0.475\textwidth]{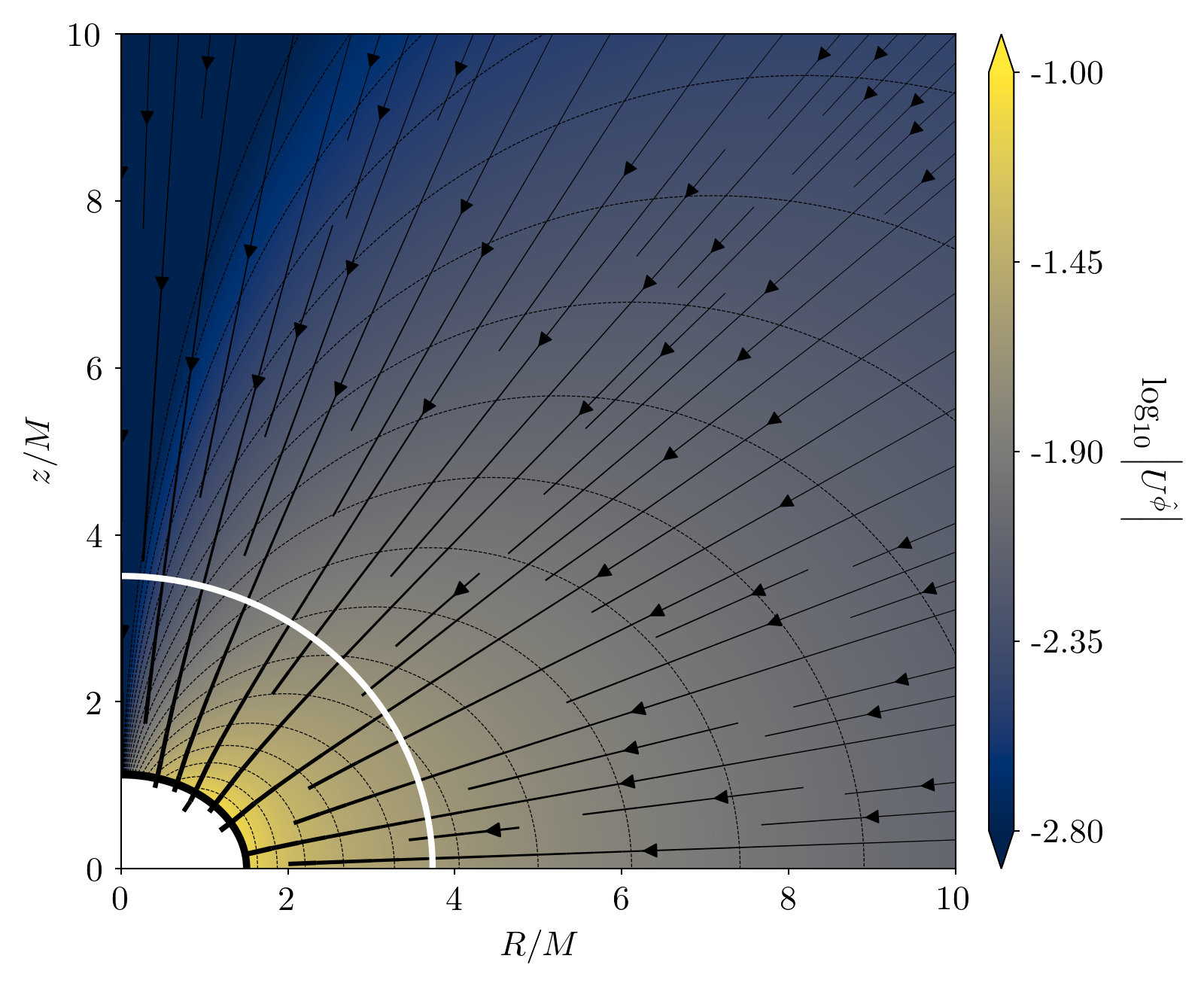}
   \caption{Isocontour plots of the steady-state of a simulation of the
   spherical accretion problem onto a rotating black hole with $a/M=0.99$,
   for a gas obeying the relativistic EoS and $\Theta_\infty = 0.1$.
   The figures show the normalised rest-mass density $\rho/\rho_\8$ at the
   equatorial plane (top-left) and the spatial orthonormal components of the
   four-velocity ($U^{\hat{r}}$ [top-right], $U^{\hat{\theta}}$ [bottom-left]
   and $U^{\hat{\phi}}$ [bottom right]) projected on the $R-z$ plane, where
   $R = \sqrt{r^2 + a^2} \, \sin \theta$ and $z = r \cos \theta$. The black
   solid arrows show the fluid streamlines, whereas the black dashed lines
   the isocontour levels. The white solid line shows the location of the
   sonic surface, see Figure~\ref{fig:sonic} for further details. The outer
   boundary in this simulation is $\mathcal{R}_{\rm out} \approx 147 \, M$.}
   \label{fig:2d}
\end{figure*}

In order to study the dependence of the spherical accretion solution on
the spin parameter, we perform a series of simulations varying the value
of $a$. For these runs, we also consider three values of the asymptotic
temperature corresponding to the non-relativistic, intermediate, and
ultra-relativistic regimes. In Figure~\ref{fig:mdot_vs_a} we show our analysis
of this dependence adopting two fluid models: the stiff fluid ($\gamma = 2$,
top panel) and the relativistic EoS (bottom panel). The former case allows us
to study the behaviour of the simulations for an extreme adiabatic index (for
which the spin effects are more noticeable), while the latter constitutes
a more realistic EoS. As in Figure~\ref{fig:mdota99}, the $\gamma = 2$
and $\Theta_\8 = 10^2$ case matches the analytic PST solution, providing
yet another code validation, but now for a wide range of spin values.

As can be seen in Figure~\ref{fig:mdot_vs_a}, the mass accretion rate decreases
as the spin parameter $a$ increases. Moreover, the dependence on $a$ becomes
more significant as higher temperatures are considered. In the case of the
stiff fluid (top panel), we find that the mass accretion rate is reduced by
up to a factor of $50\%$ for a maximally rotating black hole as compared to a
non-rotating one. On the other hand, this reduction is at most of $\sim 10\%$
in the case of the relativistic EoS (bottom panel). It is interesting to note
that all the numerical results follow a qualitatively similar dependence on
$a$ as the analytic PST solution: the accretion rate decreasing as the spin
parameter increases.\footnote{In this regard, it is interesting to mention
the recent work by~\cite{cieslik2020} who study the spherical accretion of
a Vlasov gas onto a (charged) Reissner-Nordstr\"om black hole which is often
considered as a simpler model for the Kerr spacetime since it shares many of
its qualitative properties. In this model, the charge parameter plays the role
of the spin parameter, and similar to our findings, the authors of that study
find that the mass accretion rate decreases as the charge parameter increases.}

\subsubsection{Global effect of the spin}

The dependence on the spin parameter has been studied so far by considering
only its effect on the mass accretion rate. This is important since one
of the most relevant results of any accretion model is the associated mass
growth of the central object. Nevertheless, it is also of interest to study
the overall morphology of the resulting accretion flow in order to understand
the global effect of the spin.

To study the effect of the spin on the velocity field, as well as on the
rest-mass density profile, we take as a representative example one of
the simulations discussed in Section~\ref{S3.4.2}, namely that of a fluid
obeying the relativistic EoS, with an asymptotic temperature $\Theta_\8 =
0.1$, and a spin parameter $a/M = 0.99$.

In Figure~\ref{fig:2d} we show the steady-state rest-mass density
$\rho/\rho_\8$ at the equatorial plane and the spatial components of the
velocity field $U^{\hat{r}}$, $U^{\hat{\theta}}$, $U^{\hat{\phi}}$ (measured
in an orthonormal reference frame, see Appendix~\ref{App:Frame}). The
solid black arrows show the fluid streamlines and the solid white line
represents the location of the sonic surface (see Appendix~\ref{App:sonic}
for its invariant determination). Note that the azimuthal flow shown in the
rest-mass density and in the $U^{\hat{\phi}}$ field, is due exclusively to
the frame dragging of the black hole.

\begin{figure}
   \centering
   \includegraphics[width=0.47\textwidth]{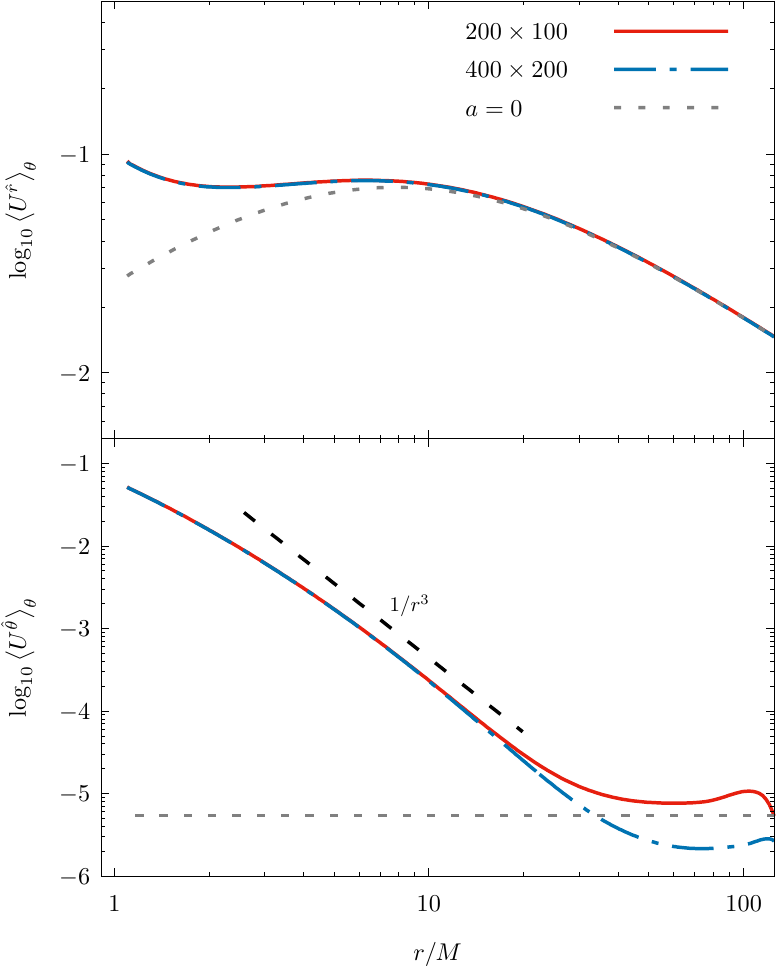}
   \caption{Solutions of the angular averaged values of $U^{\hat{r}}$ (top
   panel) and $U^{\hat{\theta}}$ (bottom panel) as a function of $r$, for
   $a/M=0.99$ and for two different resolutions. The parameters used in this
   plots are $\Theta_\8 = 0.1$ and the relativistic EoS. The grey dotted lines
   show the Michel's $a=0$ solution in the top panel, and the average numerical
   error in the polar velocity in the bottom panel. Note that the difference
   between the two resolutions in the bottom panel for large radii is of the
   same order as the average numerical error. The black dashed represents
   the approximate behaviour of $\langle U^{\hat{\theta}}\rangle_{\theta}$.}
   \label{fig:comp}
\end{figure}

As can be seen from Figure~\ref{fig:2d}, the polar component $U^{\hat{\theta}}$
exhibits a quadrupolar-like morphology, which is in contrast to the
non-rotating case where $U^{\hat{\theta}} = 0$. This is interesting since
in the PST solution this component of the four-velocity is exactly zero,
independently of the value of the spin parameter (see equation~\ref{es.12}).
On the other hand, the isocontours for $U^{\hat r}$ depart from spherical
symmetry close to the event horizon, in particular inside the sonic
surface. However, apart from the inspiraling effect due to the frame
dragging, the fluid streamlines do not deviate significantly from those of
the spherically symmetric inflow.

In order to analyse the behaviour of the fluid velocity, we compute the
latitudinal average at each radius, defined as,
\begin{equation}
   \left\langle U^{\hat{i}} \right\rangle_\theta = \frac{\displaystyle
   \int_0^{\pi/2} U^{\hat{i}} \sqrt{-g} \, \mathrm{d} \theta}{\displaystyle
   \int_0^{\pi/2} \sqrt{-g} \, \mathrm{d} \theta}.
\end{equation}
In Figure~\ref{fig:comp} we show this average for $U^{\hat{r}}$ (upper-panel)
and $U^{\hat{\theta}}$ (lower-panel) as a function of $r$ for $a/M = 0.99$
and $\Theta_\8 = 0.1$. We also use two different numerical resolutions in
order to show that our results are robust with respect to the grid size.
The black dotted line represents the non-rotating Michel solution in the
radial velocity case (top-panel), and the average numerical error that
we obtain from our simulations in the polar velocity (bottom-panel).
We find that the average of $U^{\hat{r}}$ is larger for a rotating
black hole than for a non-rotating one. Also, in the rotating case, the
average of $U^{\hat{\theta}}$ is comparable in size to $U^{\hat{r}}$ at the
horizon and decreases approximately as $1/r^3$ for $r > r_+$. The fact that
$U^{\hat{\theta}}$ is different from zero is relevant in view of previous
work~\citep[see][]{shapiro1974,zanotti2005}, which discuss spherical accretion
models in Kerr spacetime based on the assumption $U^\theta = 0$.

\begin{figure}
   \centering
   \includegraphics[width=0.475\textwidth]{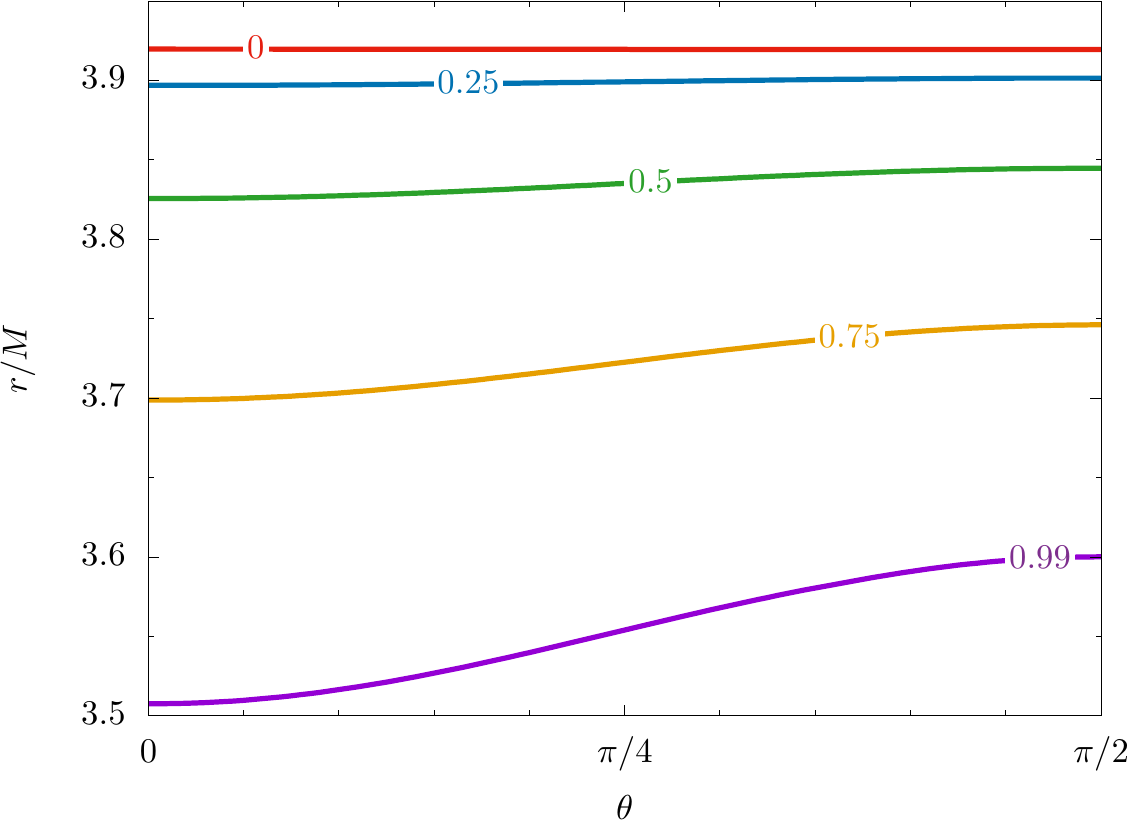}
   \caption{Resulting sonic surface as a function of the polar angle, for
   different values of the spin parameter $a/M$ (as indicated by the label
   on top of each curve). All cases correspond to the relativistic EoS
   with $\Theta_\8=0.1$. This figure shows that, for a rotating black hole,
   the sonic surface contracts to smaller radii and ceases to be characterised
   by a constant radius.}
   \label{fig:sonic}
\end{figure}

Finally, we explore in more detail the effect of the black hole
spin on the sonic surface. As shown in Appendix~\ref{App:sonic}, this
surface can be defined in an invariant way as those points at which
the magnitude of the three-velocity as measured by zero angular momentum
observers \citep[ZAMOs,][]{bardeen1970} equals the local sound speed.
In Figure~\ref{fig:sonic} we show the shape of the sonic surface by plotting
the sonic radius as a function of the polar angle, for different values of
the spin parameter. As can be seen from this figure, for the non-rotating
case the sonic surface corresponds to the sphere $r_s = \co$, as expected
since in this case the ZAMOs reduce to static observers. As the spin
parameter increases, the sonic surface contracts unevenly giving rise to a
slightly oblate shape in the $R-z$ plane. This flattening at the poles is
more significant as $a/M \to 1$. For the maximum value explored in this work
($a/M = 0.99$), the equator-to-poles difference in radii is of around $5\%$.

\section{Summary and conclusions}
\label{sec:summary}

In this work we have studied the spherical accretion problem from the
non-relativistic regime to the ultra-relativistic one, for both rotating
and non-rotating black holes. We have focused on steady-state solutions for a
perfect fluid obeying an ideal gas EoS and parametrised its thermodynamic state
far away from the black hole using the dimensionless temperature $\Theta_\8
= P_\8/(\rho_\8 c^2)$. We have also assumed that the gravitational field
is dominated by the black hole, such that the fluid's self-gravity can be
neglected. We first revisited the analytic solutions of~\citet{bondi1952}
and~\citet{michel1972}, and provided a quantitative comparison between
them. Next, we extended Michel's solution to the case of an ideal gas obeying
a relativistic EoS~\citep{juttner1911,synge1957}. Finally, we studied the
spherical accretion problem in the case of a rotating black hole, first by
writing the exact ultra-relativistic, stiff solution~\citep{petrich1988} in
the spherically symmetric case and then by performing general relativistic
hydrodynamic simulations of a general perfect fluid.

Concerning the comparison between the Bondi and Michel solutions, we have
shown rigorously that Michel's solution reduces to the Bondi one when the
non-relativistic limit is considered ($\Theta_\8 \ll 1$) and when $\gamma
\leq 5/3$, as expected. Importantly, when $\gamma > 5/3$, the obtained global
solution is intrinsically relativistic, even for non-relativistic asymptotic
temperatures, in accordance with~\cite{richards2021a}. Additionally, we derived
appropriate analytic expressions for the mass accretion rate for the Michel
solution in the ultra-relativistic limit ($\Theta_\infty \gg 1)$. Moreover,
within this limit and for a stiff EoS ($\gamma = 2$), we have shown that the
resulting mass accretion rate coincides exactly with the result obtained
by~\citet{petrich1988}. Furthermore, we have shown that in the isothermal
limit, in which $\gamma \to 1$, the entire accretion flow can be described
in a Newtonian way, i.e. the Michel solution reduces to the Bondi one for
all asymptotic temperatures $\Theta_\infty$.
   
Regarding the relativistic regime, we have found that the difference between
the mass accretion rates as obtained in the Bondi and Michel solutions grows
arbitrarily as the asymptotic temperature increases. The reason behind this
relies in the fact that, at ultra-relativistic temperatures ($\Theta_\infty
\gg 1)$, the Michel mass accretion rate reaches a minimum constant value,
whereas the Bondi one decreases without limit (Figure~\ref{fig:MmMb}). The
discrepancy between these two values is already noticeable (of order
one) for $\Theta_\infty \sim 0.1$. Moreover, we have extended the Michel
solution by considering the relativistic EoS of an ideal, monoatomic
gas~\citep{juttner1911,synge1957}, which is a more accurate description for
a perfect fluid in this regime (Figure~\ref{f_synge}).

We have also extended, by means of numerical simulations, the Michel solution
to the case of a rotating Kerr black hole. The main purpose of this numerical
exploration was to analyse the effect of the black hole spin on the mass
accretion rate and the flow morphology. We ran a series of 2.5D general
relativistic hydrodynamic simulations varying different parameters including
the asymptotic temperature $\Theta_\8$, the gas EoS, and the spin parameter
of the black hole. We have validated our results by comparing them with
the known analytic solutions, as well as by performing a series of careful
resolution and domain-size convergence tests.

The numerical results show that the influence of the black hole's rotation
is only larger than a few percent in the relativistic regime $(\Theta_\8
\gtrsim 0.1)$ or for $\gamma > 5/3$ (Figure~\ref{fig:mdota99}). As the spin
parameter increases, the mass accretion rate decreases as compared with the
non-rotating case. This effect is stronger for larger values of $\Theta_\8$
and $\gamma$. Nevertheless, even in the most extreme case ($\Theta_\infty \gg
1$ and $\gamma = 2$), the reduction in the accretion rate is no larger than
$50\%$ (Figure~\ref{fig:mdot_vs_a}). The simulations in this work allowed us
to study the morphology of the fluid's density profile and velocity field near
the event horizon, showing in the latter a behaviour considerably different
from the non-rotating black hole case, even for mildly relativistic
temperatures. We have shown that the black hole rotation induces an azimuthal
velocity component (entirely due to relativistic frame-dragging), a non-zero
polar angular velocity component, as well as a non-spherically symmetric
radial component (Figures~\ref{fig:2d} and \ref{fig:comp}). Furthermore,
the sonic surface ceases to be characterised by a constant radial coordinate
(Figure~\ref{fig:sonic}).

Our results imply that the relativistic features of a black hole can
be safely neglected when considering the spherical accretion of a fluid
with a non-relativistic asymptotic temperature ($\Theta_\infty \ll 1$)
and $\gamma\le5/3$. However, this is not true for relativistic and
ultra-relativistic values of the asymptotic temperature ($\Theta_\infty
\gtrsim 0.1$). In this regime, a proper relativistic description must be
used in order to compute the mass accretion rate, as the Bondi and Michel
solutions lead to completely different values. On the other hand,
the black hole's rotation, even in the ultra-relativistic case and for a
close-to-maximally rotating black hole, does not change the resulting mass
accretion rate by more than 50\% (for $\gamma \leq 2$) with respect to the
non-rotating case. For a more realistic EoS ($\gamma = 4/3$) this change
is even smaller and lies below 10\%. Thus, it is safe to neglect the black
hole spin when considering an order of magnitude estimation, but it should
be taken into account when performing a more accurate calculation.

The results presented in this work could be useful for studying spherical
accretion onto rotating and non-rotating black holes in extreme environments
where the ambient gas approaches relativistic temperatures ($\Theta_\8\sim
1$), or that are well approximated by a stiff EoS ($\gamma> 5/3$). Examples
of such scenarios might range from primordial black holes accreting during
the radiation era in the early universe evolution (especially between the
quark and lepton epochs when $10^{10}~\mathrm{K}<T<10^{15}~\mathrm{K}$)
\citep{jedamzik1997,lora2013a}, to mini black holes accreting from within
a neutron star (whose core can be modelled, as a first approximation, with
a $\gamma = 2$ polytrope) \citep{capela2013,genolini2020}.

\section*{Data availability}

All of the simulations presented in this work can be reproduced using the 
``Spherical accretion'' setup of the {\sc aztekas} code that can be found
on the Github repository (\url{https://github.com/aztekas-code/aztekas-main}). 
Any further data underlying this paper will be shared upon request to the 
corresponding author.

\section*{Acknowledgements}

We thank John Miller for insightful discussions and critical comments
on the manuscript. The authors also acknowledge useful comments from an
anonymous referee. This work was partially supported by CONACyT Ciencia
de Frontera Project No. 376127 ``Sombras, lentes y ondas gravitatorias
generadas por objetos compactos astrof\'\i sicos" and by a CIC grant
to Universidad Michoacana. The authors acknowledge the support from the
Miztli-UNAM supercomputer (project LANCAD-UNAM-DGTIC-406). AAO acknowledge
support from CONACyT scholarship (No. 788898).

\appendix
\section{Limits of the Michel solution}
\label{app:michel}

In this appendix we make a few remarks regarding the following two limits of
the Michel solution: the isothermal limit for which the adiabatic index
$\gamma\to1$ and the non-relativistic limit for which the asymptotic
temperature is $\Theta_\infty \ll 1$.


\vspace{8pt}
\noindent {\em (i) Isothermal limit}
\vspace{8pt}

In the limit when $\gamma\to 1$, we show that the Michel solution approaches
the Newtonian (Bondi) flow solution with an EoS given by \eq{eb.6}. To this
end, we first use the cubic \eq{poly} and find, for small values of $\delta :=
\gamma - 1 > 0$,
\begin{subequations}
\begin{gather}
   \frac{h_s}{h_\8} = 1 + \frac{3}{2}\delta - \left( \frac{9}{8}
   + \frac{3}{2\Theta_\8} \right)\delta^2 + {\cal O}(\delta^3),\\
   \mathcal{C}^2_\8 = \delta\left[ 1 - \frac{\delta}{\Theta_\8} + {\cal
   O}(\delta^2) \right], \label{Eq:cinf}\\
   \frac{\mathcal{C}_s}{\mathcal{C}_\8} = 1 + {\cal O}(\delta)^2,
\end{gather}
\end{subequations}
from which
\begin{equation}
   \frac{\dot{M}_M}{4\pi M^2 \rho_\8 \mathcal{C}^{-3}_\8}\to \frac{1}{4}
   e^{3/2},
   \label{Eq:Limitgam1}
\end{equation}
which coincides with the Bondi solution in \eq{bm}.

Next, we introduce the dimensionless quantities
\begin{equation}
   x := \frac{r}{M} \mathcal{C}^2_\8,\quad
   z := \frac{\rho}{\rho_\8},\quad
   \nu := \frac{u}{c},\quad
   \lambda := \frac{\dot{M}_M}{4\pi M^2}\frac{\mathcal{C}^3_\8}{\rho_\8}.
   \label{Eq:DimLess}
\end{equation}
in terms of which eqs.~(\ref{em.3},\ref{em.4}) can be rewritten as
\begin{subequations}
\begin{gather}
   x^2\nu z^{\frac{\gamma+1}{2}} = \lambda\left( \frac{h}{h_\8} \right)^{1/2},
   \label{Eq:Limit1}\\
   -\frac{2}{x} + \frac{h_\8}{h} z^{\gamma-1}\nu^2 =
   \frac{1}{\mathcal{C}^2_\8}\left[ \left( \frac{h_\8}{h}\right)^2 - 1
   \right]. \label{Eq:Limit2}
\end{gather}
\end{subequations}
For small values of $\delta$, one finds, using $h = 1 +
\gamma\rho^\delta/\delta$,
\begin{equation}
   \frac{h}{h_\8} = 1 + \delta\log(z) + {\cal O}(\delta^2).
\end{equation}
Introduced into \eqs{Eq:Limit1},\eqref{Eq:Limit2}, using \eq{Eq:cinf} and
taking the limit $\delta\to 0$ yields (assuming that $x$, $z$ and $\nu$
have finite values in this limit)
\begin{equation}
   x^2\nu z = \lambda,\qquad -\frac{1}{x} + \frac{1}{2}\nu^2 = -\log z,
\end{equation}
which agrees precisely with the Newtonian equations~(\ref{eb.3},\ref{eb.4})
with the EoS~(\ref{eb.6}), for $x$, $z$ and $\lambda$ defined as
in Eq.~(\ref{Eq:DimLess}) and $\nu = v/\mathcal{C}_\8$ (note that
$c/\mathcal{C}_\8\to 1$ in the limit $\delta\to 0$). Taking into account the
limit~(\ref{Eq:Limitgam1}) this yields the transonic flow solution discussed in
subsection~\ref{SubSec:Bondi} which has been shown in Ref.~\cite{chaverra2016a}
to be the correct $\gamma\to 1$ limit of the Bondi flow.


\vspace{8pt}
\noindent {\em (ii) Non-relativistic limit}
\vspace{8pt}

In the low-temperature limit $\Theta_\8\to 0$ one has $h_\8\to 1$, and in
this limit \eq{poly} has two positive roots
\begin{equation}
   h_s = 1,\qquad
   h_s = \frac{1}{2}\left( \sqrt{12\gamma - 11} - 1 \right),
   \label{Eq:roots}
\end{equation}
the third one being negative and hence unphysical. For $\gamma < 5/3$ the
second positive root is smaller than one, and hence unphysical as well and
the correct limit is $h_s = 1$. Computing the first-order correction in
$\Theta_\8$ one finds
\begin{equation}
   \frac{h_s}{h_\8} = 1 + \frac{3\gamma}{5 - 3\gamma}\Theta_\8 +
   {\cal O}(\Theta_\8)^2,
\end{equation}
from which
\begin{equation}
   \frac{\mathcal{C}_s^2}{\mathcal{C}^2_\8} = \frac{2}{5 - 3\gamma} +
   {\cal O}(\Theta_\8),
\end{equation}
and substituting into \eq{dMM} it follows that $\dot{M}_M\to \dot{M}_B$ when
$\Theta_\8\to 0$ and $\gamma < 5/3$. When $\gamma > 5/3$ it turns out the
correct root is the second one in \eq{Eq:roots}, see \cite{chaverra2016b},
and the corresponding squared sound
speed and radius at the sonic point are
\begin{equation}
   \mathcal{C}_s^2 = \frac{1}{3}(h_s ^2 - 1) > 0, \qquad r_s = \frac{3M}{2}\frac{h_s^2}{h_s^2-1}.
\end{equation}
It follows from \eq{dMM} that
\begin{equation}
   \dot{M}_M\to \pi h_s^{\frac{3\gamma-2}{\gamma-1}}
   \mathcal{C}_s^{\frac{5-3\gamma}{\gamma-1}} M^2\rho_\8\,
   \mathcal{C}^{-\frac{2}{\gamma-1}}_\8,
\end{equation}
and the mass accretion rate decays slower than $\mathcal{C}^{-3}_\8$. Note
that \mbox{$h_s > 1$} and $\mathcal{C}_s > 0$ imply that the flow does not
lie in the Newtonian regime close to the sonic point when $\gamma > 5/3$.

\section{Orthonormal frame adapted to the Kerr-type coordinates}
\label{App:Frame}

The orthonormal frame adapted to the constant time slices in the Kerr-type
coordinates $(t,\phi,r,\theta)$ used in this article is given by
\begin{subequations}
   \begin{align}
      e_{\hat{t}} &= \sqrt{1 + \frac{2M r}{\varrho^2}} \left(
      \frac{\partial}{\partial t} -\frac{2Mr}{\varrho^2 +
      2Mr}\frac{\partial}{\partial r} \right),
      \label{Eq:ehatt}\\
      e_{\hat{r}} &= \frac{1}{\sqrt{1 + \frac{2M
      r}{\varrho^2}}}\frac{\partial}{\partial r},
      \label{Eq:ehatr}\\
      e_{\hat{\theta}} &= \frac{1}{\varrho} \frac{\partial}{\partial \theta},
      \label{Eq:ehattheta}\\
      e_{\hat{\phi}} &= \frac{1}{\varrho\sin\theta}\left(
      \frac{\partial}{\partial \phi} + a\sin^2\theta \frac{\partial}{\partial
      r} \right),
      \label{Eq:ehatphi}\\
   \end{align}
\end{subequations}
and it is well-defined for all $r > 0$. The corresponding components of the
four-velocity vector field, such that
\begin{equation}
   U^\mu\frac{\partial}{\partial x^\mu} = U^{\hat{t}} e_{\hat{t}} + U^{\hat{r}}
   e_{\hat{r}} + U^{\hat{\theta}}
   e_{\hat{\theta}} + U^{\hat{\phi}} e_{\hat{\phi}}
\end{equation}
are given by
\begin{subequations}
   \begin{align}
      U^{\hat{t}} &= \frac{1}{\sqrt{1 + \frac{2M r}{\varrho^2}}} U^t,
      \label{eq:orthot}\\
      U^{\hat{r}} &=  \sqrt{1 + \frac{2M r}{\varrho^2}} \left( U^r +
      \frac{2Mr}{\varrho^2 + 2Mr} U^t - a\sin^2\theta U^\phi \right),
      \label{eq:orthor}\\
      U^{\hat{\theta}} &= \varrho U^\theta, 
      \label{eq:orthotheta}\\
      U^{\hat{\phi}} &= \varrho\sin\theta U^\phi.
      \label{eq:orthophi}
   \end{align}
\end{subequations}

\section{Invariant determination of the sonic surface}
\label{App:sonic}

In relativistic fluids, it is not immediately obvious how to determine the
sonic surfaces, that is, the boundary separating the events at which the
flow is subsonic from those at which it is supersonic. Indeed, the fluid's
sound speed $\mathcal{C}$ is a scalar, while the velocity $U^\mu$ of the
fluid is a four-vector. One could consider instead of $U^\mu$ the magnitude
of the three-velocity $V$ with respect to a specific family of observers
and define the sonic surface by those events for which $V = \mathcal{C}$,
but this definition would clearly be observer-dependent.

A definition which does provide an invariant characterization of the sonic
surface is based on the sonic metric,
\begin{equation}
   \mathfrak{G}_{\mu\nu} := \frac{\rho}{h}\frac{1}{\mathcal{C}}\left[
   g_{\mu\nu} + \left(1 - \mathcal{C}^2 \right) U_\mu U_\nu \right],
   \label{Eq:SonicMetric}
\end{equation}
first introduced by~\cite{moncrief1980}, for the purpose of analysing
the propagation linearised, acoustic perturbations of an isentropic,
vorticity-free flow on a background spacetime with metric $g_{\mu\nu}$. The
sonic metric~(\ref{Eq:SonicMetric}) is a Lorentzian metric whose set of null
vectors at a given spacetime event $e$ form a cone (the sound cone) that can be
shown to lie inside the light cone at $e$ provided $\mathcal{C}^2 < 1$. Another
useful property of the sonic metric is that it inherits the symmetries of
the spacetime and the flow configuration: if $X$ is a Killing vector field,
such that the Lie derivative $\pounds_X$ of $g_{\mu\nu}$, $U^\mu$, $\rho$
and $h$ vanish, then it follows that $\pounds_X \mathfrak{G}_{\mu\nu} = 0$,
that is, the sonic metric is invariant with respect to $X$.

For the solutions described in this article, where both the spacetime
metric and the flow are steady-state and axisymmetric, it follows that
Eq.~(\ref{Eq:SonicMetric}) describes a steady-state and axisymmetric geometry
which is asymptotically flat since the flow's four-velocity is constant
at infinity. A sonic surface corresponds to the ``event horizon'' of this
geometry, that is, the surface which separates those events that can send an
acoustic signal to infinity from those that cannot. Due to the aforementioned
symmetries of the sonic geometry, this surface must be a Killing horizon,
i.e. a null surface of the form \citep[see, e.g.][]{heusler1996}
\begin{equation}
   {\cal H} := \{x : \mathfrak{G}_{\mu\nu}(x) X^\mu X^\nu = 0 \},
\end{equation}
whose normal vector,
\begin{equation}
   X^\mu = \delta^\mu{}_t + \Omega_{\cal H}\delta^\mu{}_\phi,
\end{equation}
is a superposition of the Killing vector fields of the Kerr metric, where
here the constant $\Omega_{\cal H}$ describes the angular velocity of the
horizon. The requirement of ${\cal H}$ being a null surface (with respect
to the sonic metric) with normal $X^\mu$ implies the condition
\begin{equation}
   \nabla^\alpha\left[ \mathfrak{G}_{\mu\nu}(x) X^\mu X^\nu \right] =
   -2\kappa X^\alpha,
   \label{Eq:NullCondition}
\end{equation}
the proportionality factor $\kappa$ describing the ``surface gravity"
associated with the horizon. Assuming a regular horizon, such that $\kappa
\neq 0$, the four equations~(\ref{Eq:NullCondition}) imply
\begin{subequations}
   \begin{align}
      &\mathfrak{G}_{tt} + \Omega_{\cal H} \mathfrak{G}_{t\phi} = 0,
      \label{Eq:Det1}\\
      &\mathfrak{G}_{t\phi} + \Omega_{\cal H} \mathfrak{G}_{\phi\phi} = 0,
      \label{Eq:Det2}\\
      &\mathfrak{G}_{tr} + \Omega_{\cal H} \mathfrak{G}_{\phi r} = 
      -\frac{1}{2\kappa}\frac{\partial N}{\partial r},
      \label{Eq:kappa1}\\
      &\mathfrak{G}_{t\theta} + \Omega_{\cal H} \mathfrak{G}_{\phi\theta} = 
      -\frac{1}{2\kappa}\frac{\partial N}{\partial \theta}
      \label{Eq:kappa2}
   \end{align}
\end{subequations}
with $N := \mathfrak{G}_{\mu\nu} X^\mu X^\nu = \mathfrak{G}_{tt}
+ 2\Omega_{\cal H} \mathfrak{G}_{t\phi} + \Omega_{\cal
H}^2\mathfrak{G}_{\phi\phi}$. Note that the first two
conditions~(\ref{Eq:Det1},\ref{Eq:Det2}) imply that $X^\mu$ is null on
${\cal H}$, i.e. $N = 0$, as required. They determine the location of the
sonic surface ${\cal H}$ through the requirement that the determinant of
the $2\times 2$ matrix $(\mathfrak{G}_{ab})_{a,b = t,\phi}$ vanishes. In
view of definition~(\ref{Eq:SonicMetric}) this yields
\begin{equation}
   \det\left[ g_{ab} + (1-\mathcal{C}^2) U_a U_b \right] = 0.
   \label{Eq:Det}
\end{equation}
In turn, either Eq.~(\ref{Eq:kappa1}) or Eq.~(\ref{Eq:kappa2}) can be used
to determine the surface gravity $\kappa$, but this will not be needed
here.\footnote{Note that the condition $N = 0$ on ${\cal H}$ implies
that Eq.~(\ref{Eq:NullCondition}), when contracted with a tangent vector
to ${\cal H}$ is automatically satisfied, such that only one of the two
equations~(\ref{Eq:kappa1},\ref{Eq:kappa2}) needs to be considered.}

In terms of the Kerr-type coordinates $(t,\phi,r,\theta)$ used in this article,
the determinant condition~(\ref{Eq:Det}), together with the property $U_\phi =
0$ satisfied by the flow, leads to the condition
\begin{equation}
   g_{tt} - \frac{g_{t\phi^2}}{g_{\phi\phi}} + (1-\mathcal{C}^2) U_t^2 = 0,
   \label{Eq:SonicHorizon}
\end{equation}
which yields an implicit relation between $r$ and $\theta$. This condition
acquires a much clearer interpretation when rewriting it in terms of the
flow's Lorentz factor $\Gamma$ measured by a ZAMO, which gives
\begin{equation}
   (1-\mathcal{C}^2)\Gamma^2 = 1,
\end{equation}
i.e. the sonic surface is determined by those events for which the
flow, as measured by ZAMOs, changes from sub- to supersonic. From
Eq.~(\ref{Eq:Det2}) and $U_\phi = 0$ it also follows that $\Omega_{\cal H}
= -g_{t\phi}/g_{\phi\phi} = \Omega_{\rm ZAMO}$, i.e. the angular velocity of
the sonic horizon is equal to the angular velocity of the ZAMO at ${\cal H}$.


\bibliographystyle{mnras}
\bibliography{references}

\label{lastpage}

\end{document}